\documentclass[preprintnumbers,
superscriptaddress,
prd,nofootinbib,floatfix,12pt
]{revtex4}

\usepackage[normalem]{ulem}
\usepackage{graphicx,amssymb,amsmath}
\usepackage{hyperref}
\usepackage{color}
\usepackage{setspace}

\renewcommand{\arraystretch}{1.3}

\def\beq{\begin{equation}}
\def\eeq{\end{equation}}

\def\bea{\begin{eqnarray}}
\def\eea{\end{eqnarray}}
\def\bei{\begin{itemize}}
\def\eei{\end{itemize}}
\def\bee{\begin{enumerate}}
\def\eee{\end{enumerate}}
\def\bmat{\begin{matrix}}
\def\emat{\end{matrix}}
\def\={\,=\,}
\def\+{\,+\,}
\def\-{\,-\,}

\def\MET{E_T^{\textrm{miss}} }

\newcommand{\Fig}[1]{Fig.~\ref{#1}}
\newcommand{\Eq}[1]{Eq.~(\ref{#1})}
\newcommand{\Sec}[1]{Sec.~\ref{#1}}

\begin{document}
\title{
Prospects for Electroweakino Discovery \\
at a 100\,TeV Hadron Collider
}

\author{\vspace{2mm} Stefania Gori}
\email{sgori@perimeterinstitute.ca}
\affiliation{Perimeter Institute for Theoretical Physics, 31 Caroline St. N, Waterloo, Ontario, Canada N2L 2Y5.}

\author{Sunghoon Jung}
\email{nejsh21@kias.re.kr}
\affiliation{Korea Institute for Advanced Study, Seoul 130-722, Korea}

\author{Lian-Tao Wang}
\email{liantaow@uchicago.edu}
\affiliation{Enrico Fermi Institute, University of Chicago, Chicago, IL 60637}
\affiliation{Kavli Institute for Cosmological Physics, University of Chicago, Chicago, IL 60637}

\author{James D.~Wells}
\email{jwells@umich.edu}
\affiliation{Michigan Center for Theoretical Physics, Department of Physics, University of Michigan, Ann Arbor, MI 48109}

\begin{abstract} \vspace{3mm} \baselineskip=16pt

We investigate the prospects of discovering split Supersymmetry at a future 100 TeV proton-proton collider through the direct production of electroweakino next-to-lightest-supersymmetric-particles (NLSPs).
 We focus on signatures with multi-lepton and missing energy: $3\ell$, opposite-sign dileptons and same-sign dileptons. We perform a comprehensive study of different electroweakino spectra. A 100 TeV collider with $3000/$fb data is expected to exclude Higgsino thermal dark matter candidates with $m_{\rm{LSP}}\sim 1 $ TeV if Wino NLSPs are lighter than about 3.2 TeV. The $3\ell$ search usually offers the highest mass reach, which varies in the range of (2--4) TeV depending on scenarios. In particular, scenarios with light Higgsinos have generically simplified parameter dependences.
We also demonstrate that, at a 100 TeV collider, lepton collimation becomes a crucial issue for NLSPs heavier than about 2.5 TeV.
We finally compare our results with the discovery prospects of gluino pair productions and deduce which SUSY breaking model can be discovered first by electroweakino searches.

\end{abstract}

\preprint{KIAS-P14050}

\maketitle

\newpage

\baselineskip=15pt
\tableofcontents

\vfill\eject

%%%%%%%%%%%%%%%%%%%%%%
\baselineskip=18pt

\section{Introduction}

The lack of discovery of Supersymmetry (SUSY) at the first run of the LHC started to place some tension on natural SUSY. Even though it is premature to abandon the idea of a natural spectrum, 
an attractive scenario is split SUSY~\cite{Wells:2003tf,ArkaniHamed:2004fb,Giudice:2004tc,Wells:2004di}. 
We refer to, e.g., Refs.~\cite{Randall:1998uk,Giudice:1998xp,Arvanitaki:2012ps,McKeen:2013dma,ArkaniHamed:2012gw,Kahn:2013pfa,Bhattacherjee:2012ed} for developments along this line. 

In these models, gauginos and higgsinos are the lightest SUSY particles and provide most important collider search channels as SUSY scalars are much heavier. Collider searches of gluino pair production usually lead to the easiest discovery if gluinos are not much heavier than other gauginos and higgsinos. According to Refs.~\cite{Cohen:2013xda,Jung:2013zya},  gluinos up to about 11 TeV can be discovered at a 100 TeV proton-proton collider with 3 ab$^{-1}$ data. If gluinos are heavier, electroweakinos\footnote{We use the term ``electroweakinos" throughout to refer to Winos, Binos and Higgsinos.} can be the best channel to discover split SUSY. In any case, electroweakino studies are essential for precision measurements of the superpartner mass spectrum. Gluino-focused studies are not enough in this regard as $M_{eff}$ from gluino pair decays is not very sensitive to the electroweakino mass spectrum~\cite{Jung:2013zya}. 

Electroweakino searches can also probe the WIMP (weakly interacting massive particle) nature of neutralino lightest superparticles (LSPs). Either Higgsinos, Winos or well-tempered neutralinos can serve as thermal relic cold dark matter (DM) candidates with full relic abundance as needed to satisfy cosmological data~\cite{ArkaniHamed:2006mb}. 1 TeV and 3.1 TeV are the masses of potential Higgsino and Wino full thermal DM~\cite{Fan:2013faa}. Testing the split parameter space up to these masses is both an important mission and a useful goal of a future collider. Direct LSP collider searches are the most model independent tests of the scenario. According to dedicated studies in Refs.~\cite{Low:2014cba,Cirelli:2014dsa}, Wino DM can plausibly be probed at a 100 TeV collider, but probing Higgsino DM through those searches will be unlikely.

In this paper, we study the 100 TeV proton-proton collider prospects of NLSP electroweakino searches in multi-lepton final states. In particular, we will discuss the potential of probing Higgsino dark matter from the pair production of Wino NLSPs and Wino dark matter from the pair production of Higgsinos NLSPs. Finally, we will compare the search capabilities of these channels to those based on direct gluino production and decay.

\medskip 

Meanwhile, this parameter space of SUSY, with relatively light (at most few TeV) electroweakinos and much heavier scalar superpartners, must be studied in qualitatively different ways in several aspects, compared to the previous studies of ${\cal O}(100)$ GeV SUSY at the LHC. 
As is well known, boosted phenomena and electroweak radiation phenomena become central issues at a 100\,TeV proton collider; see, e.g. Ref.~\cite{Hook:2014rka}. Moreover, more analytic approaches are possible for this higher energy environment with only electroweakinos  accessible, as a smaller number of particles and parameters are relevant to the final signatures.

The very high energy of the collisions with relatively light electroweakinos create, in fact, an environment where  the Goldstone equivalence theorem generically applies. Therefore, the various electroweakino decay branching ratios (BRs) are inherently related. Interestingly, the NLSP BRs involving Higgsinos (either as decaying mother particles or daughter particles) are greatly simplified in this parameter space. All the underlying dependences from $\tan \beta$ and from the signs of gaugino and higgsino masses essentially vanish as a result of (1) summing the effects of two \emph{indistinguishably} degenerate neutral Higgsinos to calculate what we actually \emph{observe} at the collider~\cite{Jung:2014bda} and (2) the Higgs alignment limit dictated by Higgs signal strength data~\cite{Jung:work}. 
We emphasize that these relations did not hold previously in general, especially when the  electroweakinos under consideration are light.  At the same time, they become very good approximations for TeV scale electroweakinos. Various other relations are also revealed in a similar way and analytic understanding of BRs are aided~\cite{Jung:2014bda,Han:2013kza}.

Throughout this paper, we present results obtained by full numerical computation of BRs. As already mentioned, we have model independent BRs in the scenarios with Higgsino-NLSP or -LSP. In the case of heavier Higgsinos, $\mu> M_2, M_1$, the results will be more model dependent. For this reason, we will consider several choices of parameters with heavier Higgsinos and provide analytic discussion.

The paper is organized as follows. We first introduce multi-lepton searches and our collider analysis strategy in \Sec{sec:search}. \Sec{sec:prospects} contains our main results: we provide discovery and exclusion prospects for several scenarios containing different types of NLSPs and LSPs. We also compare our results with the discovery prospects of split SUSY via gluino pair productions. We further discuss potential issues regarding detector and object measurements at a future 100 TeV hadron collider in \Sec{sec:detector}. We finally reserve \Sec{sec:concl} to our conclusions. We estimate several uncertainties involved in our analysis in the Appendix.

%%%%%%%
\section{Multi-Lepton Searches} \label{sec:search}

\subsection{Search Channels}
In split SUSY, all the scalars are much heavier than the electroweakinos and therefore electroweakinos can only decay to gauge bosons or to the 125 GeV Higgs boson. Electroweakino pairs thus decay to intermediate di-boson channels: di-vector $WZ$, $W^+ W^-$, $W^\pm W^\pm$, $ZZ$ and Higgs channels $Wh$, $Zh$, $hh$.  The di-vector channels are currently efficiently probed by multi-lepton searches~\cite{Aad:2014nua,Aad:2014vma,Khachatryan:2014qwa,Aad:2014pda,ATLAS:2013qla}: the three-lepton ($3\ell$), opposite-sign di-lepton (OSDL), same-sign di-lepton (SSDL) and four-lepton ($4\ell$) searches. 
The contributions to the multi-lepton signatures from the Higgs channels are generally subdominant; they can be dominant only if the branching ratio of the NLSP electroweakinos to the Higgs boson is close to 100\%.
 Although lepton plus jet searches, such as $2\ell + 2j$ from $WZ$~\cite{Khachatryan:2014qwa} and $\ell +b\bar{b}$ from $Wh$~\cite{Baer:2012ts,Han:2013kza}, can certainly be useful, our estimation based solely on multi-lepton searches  is expected to represent the search capacity of a 100 TeV collider reasonably well.
 
Here, we summarize the main features of multi-lepton searches.
We refer the reader to Sec.~\ref{sec:simdet} for simulation details, \Sec{sec:opt} for the variables used in our analysis and \Sec{sec:opt-cut} for sample optimized cut flows.

\bei

\item {\bf{3$\mathbf\ell$}}: 
This search mode usually leads to the best reach. Chargino-neutralino pair production has always the largest electroweakino pair production cross section leading to a $3\ell$ signal, mainly via the intermediate $WZ$ channel. The $Wh$ channel becomes important only when the NLSP neutralino has a very small branching ratio into $Z$ bosons.

In our analysis, we require exactly three isolated and separated leptons. 
The observables we optimize 
are missing transverse energy (MET) $\MET$, the transverse mass obtained using the MET and the lepton not belonging to the same flavor opposite sign (SFOS) lepton pair with invariant mass closest to the $Z$ mass, $M_T(W)$, $M_{eff}^\prime \, \equiv \, M_{eff}-p_T(\ell_1)$, the $p_T$ ratio of the second and leading leptons $p_T(\ell_2)/p_T(\ell_1)$, and the jet energy fraction $H_T({\rm jets})/M_{eff}$. Here, $M_{eff}$ ($H_T({\rm jets})$) is the scalar sum of $p_T$'s of reconstructed jets, leptons and MET (jets only). No explicit jet veto is applied. We will discuss in Sec.~\ref{sec:opt}, however, that the upper cut on the jet energy fraction is analogous to a jet veto.
  The SM $WZ/\gamma$ \footnote{In the following, we will always denote the $WZ/\gamma$ background by simply $WZ$.} production is the dominant background for most cases, while tribosons and $t\bar{t}V$ backgrounds can be relevant when the $Wh$ becomes the dominant contribution to the signal.

The latest $3\ell$ LHC8 searches
can be found in Refs.~\cite{Aad:2014nua,Khachatryan:2014qwa}. Wino NLSPs up to $\sim 350$ GeV are excluded for massless Bino LSPs in the simplified model, which assumes a $100\%$ branching ratio for $C_1^\pm N_2 \to W^\pm Z N_1 N_1$.

\item {\bf{Opposite-sign di-leptons (OSDL)}}: 

The $W^+W^-$ channel is the dominant signal contribution. 

In our analysis, we require exactly two opposite-sign leptons of any flavor and veto any events with reconstructed jets ($p_T>30$ GeV, $|\eta|<2.5$). 
The observables we optimize
 are $M_{eff}^\prime,$ missing energy fraction $\MET/M_{eff}, \, p_T(\ell_2)/p_T(\ell_1)$ and the transverse mass obtained from the two leptons and the missing energy, $M_T(\MET, \ell \ell)$. With our analysis, the SM $W^+W^-$ is the dominant background and the SM $WZ$ is also non-negligible.

The latest OSDL LHC8 searches can be found in Refs.~\cite{Aad:2014vma,Khachatryan:2014qwa}. Wino NLSPs up to $\sim 200$ GeV are excluded for massless LSPs in the simplified model, which only includes the process $C_1^\pm C_1^\mp \to W^\pm W^\mp N_1 N_1$ with assumed 100\% BR.

\item {\bf{Same-sign di-leptons (SSDL)}}: 

The $W^\pm W^\pm$ channel is the dominant signal contribution, but it is absent in some NLSP-LSP configurations, for which $WZ$ becomes the only channel contributing to this signature, if one of the three leptons is missed. Standard ATLAS and CMS searches show that, typically, this channel is the best search mode for electroweakino spectra with small mass gap between the NLSP and the LSP, for which one of the
leptons coming from the NLSP decay might be too soft to be included in the $3\ell$ analysis.

In our analysis, we require exactly two same-sign leptons of any flavor and veto any events with reconstructed jets ($p_T>30$ GeV, $|\eta|<2.5$). 
The observables we optimize are $M_{eff}^\prime, \, \MET, \, p_T(\ell_2)/p_T(\ell_1)$ and $M_T(\MET,\ell \ell)$. The SM $WZ$ is the dominant background while fake and mis-identified backgrounds can similar or larger~\cite{Khachatryan:2014qwa}, and the double parton scattering (DPS) production of $W^\pm W^\pm$ is smaller\footnote{The DPS $W^\pm W^\pm$ produces softer leptons than the $W^+W^-$ background~\cite{Gaunt:2010pi}, which make it less important for high mass searches. If it had the same kinematic distributions as the SM $W^+ W^-$, it would contribute to the SSDL search by only $\sim20\%$ of the main SM $WZ$ background.}. Muons are perhaps cleaner against the fake backgrounds~\cite{ATLAS:2012mn}, but we include both $e$ and $\mu$ with equal efficiencies\footnote{We learn from Maurizio Pierini that the resolution of high-$p_T$ muon measurements can be quite worse than that of high-$p_T$ electrons depending on the performances of calorimeters and magnet strengths.}.
The latest SSDL LHC8 searches can be found in Ref.~\cite{Khachatryan:2014qwa}. Wino NLSPs up to $\sim 130$ GeV are excluded for massless LSPs in the simplified model which only includes the process $N_2 C_1^\pm \to Wh N_1 N_1$ with assumed  100\% BR.

%%%%%%%%%%%%%%
\item {\bf{$\mathbf {4\ell}$}}: 
Exactly four leptons of any flavor can be searched for. The $ZZ$ channel is the dominant contribution.
Due to the small branching ratio of the $Z$ to leptons and to the smaller production cross section of neutral NLSPs, if compared to the associated production of a neutral and a charged NLSP, this channel is typically not a leading discovery channel. For this reason, we do not consider this signature further.

The latest $4\ell$ LHC8 searches can be found in Refs.~\cite{ATLAS:2013qla,Khachatryan:2014qwa}. Higgsinos up to $\sim 150$ GeV are excluded in the context of GMSB models with the gravitino LSP.

\eei 

\subsection{Analysis Details} \label{sec:simdet}

We model signals and backgrounds using MadGraph5~\cite{Alwall:2011uj}, interfaced with
Pythia 6.4~\cite{Sjostrand:2006za}, for parton showering. We allow up to one additional parton in the final state,
and adopt the MLM matching scheme~\cite{Mangano:2006rw} with \texttt{xqcut}=40 GeV. The generated SM background processes are di-boson ($WW$, $WZ$ and $ZZ$), tribosons and $t\bar t V$. We also check $t\bar{t}$ and $Wh$ backgrounds for some cases.
The backgrounds are generated in successively smaller phase spaces sectored by the scalar $p_T$ sum of intermediate dibosons~\cite{Cohen:2013xda,Jung:2013zya} as done for the Snowmass studies~\cite{Anderson:2013kxz}. To this end, we modify the MadGraph \texttt{cuts.f} code.
Corresponding Pythia matched rates are used for background normalizations. Summing all sectored backgrounds yields a rate similar to the next-to-leading order results predicted from MCFM~\cite{Campbell:1999ah}. As for signal rates, we multiply the leading order MadGraph results by the assumed NLO K-factor, $K=1.2$. 

Leptons are required to have $p_T>15$ GeV and $|\eta| <2.5$. To reconstruct jets, we use the anti-$k_T$ algorithm~\cite{Cacciari:2008gp} with $R=0.4$ implemented in FastJet-2.4.3~\cite{Cacciari:2011ma} on remaining particles. Jets are required to have $p_T>30$ GeV and $|\eta| < 2.5$. If a reconstructed lepton is found within $\Delta R=0.4$ of a reconstructed jet, the lepton is merged into the jet. We require leptons (both muons and electrons) to be separated by more than $\Delta R=0.05$. We cluster photons nearby a lepton within a cone of $\Delta R<0.05$, to reconstruct the lepton by taking into account QED radiation effects. Hard QED radiation is infrequent but resulting photons can carry away a non-negligible fraction of energy momentum. As shown in~\cite{Gori:2013ala}, this is especially important to reconstruct $Z$ peaks more properly. 
Lepton identification efficiencies are adapted from the current ATLAS efficiencies~\cite{Aad:2011mk,Aad:2009wy}. Typically, as our leptons are energetic, more than $95\%$ of the leptons are identified.
We refer to \Sec{sec:detector} for related discussions. We do not take into account any detector effects such as finite cell sizes and momentum smearing.

Our baseline selection requires no additional leptons: exactly $3\ell$, OSDL and SSDL for corresponding searches. A jet veto is applied for the OSDL and SSDL. Any SFOS dileptons are required to have invariant mass, mSFOS $>12$ GeV. In addition, the mSFOS closest to $m_Z$, denoted by mSFOS(Z), should be within (outside of)  30 GeV of $m_Z$ if $WZ \to 3\ell$ ($Wh \to 3\ell, W^\pm W^\mp \to 2\ell, W^\pm W^\pm \to 2\ell$) channels are searched for. For the OSDL, we additionally apply $p_T(\ell \ell)>30$ GeV, where $p_T(\ell \ell)$ is the transverse momentum of the lepton pair. Finally, we require either MET$>100$ GeV or $p_T(l_1)>100$GeV. No specific trigger is applied but we expect that devising a working trigger will not be problematic, thanks to the high energy cuts used in our analysis.

%%%
\subsection{The Variables Used in Analysis} \label{sec:opt}

The search for the high-mass and large-gap parameter space is based on high visible and invisible energy. Signal processes easily produce such high energy particles from decays of heavy mother particles. The SM backgrounds, instead, can reach high visible and invisible energies only by
hard radiations. This implies certain alignment between the boost direction  
and the final state particle momentum so that certain particles are more efficiently boosted. 

\medskip

\noindent
\underline{ $3\ell$ Search: }

\begin{figure}[t] 
\includegraphics[width=0.95\textwidth]{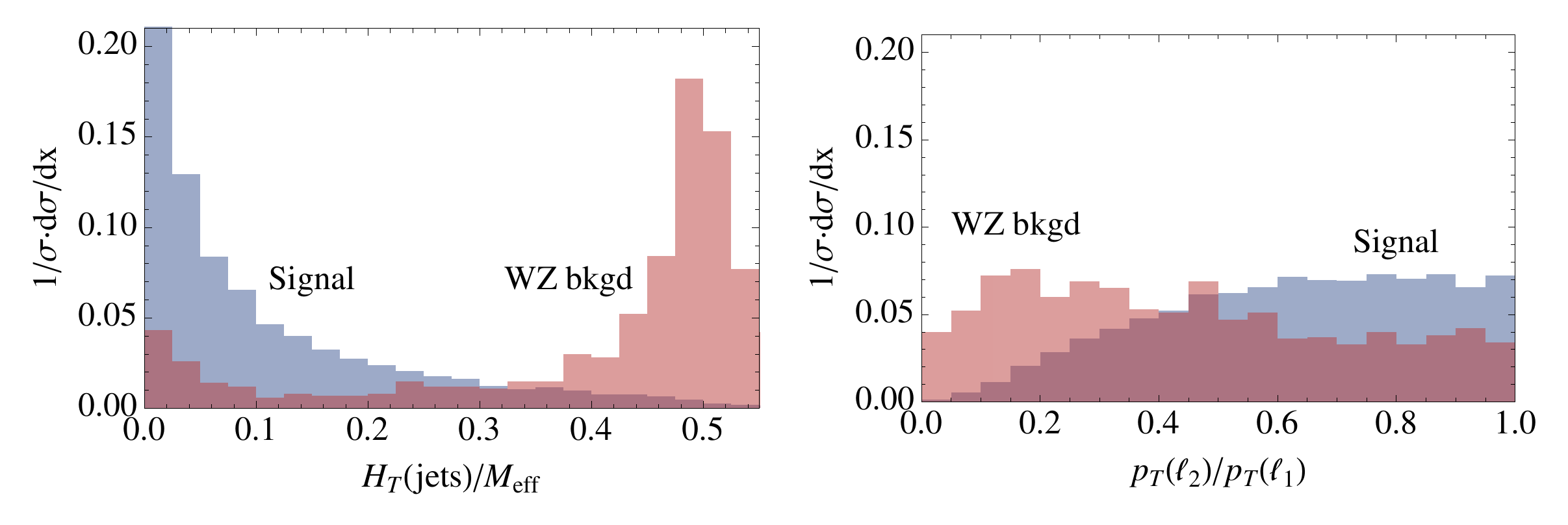}
\caption{Normalized distributions of observables optimized for the $3\ell$ search.  $H_T({\rm jets})/M_{eff}$ {(left panel)} and $p_T(\ell_2)/ p_T(\ell_1)$ {(right panel)}. $WZ$ signal events (blue) are from the benchmark with 3 TeV NLSP and massless LSP, and backgrounds events (red) are from the SM $WZ$. All the discovery cuts in Table~\ref{tab:WH-3l-cutflow} are applied  except for those on $p_T(\ell_2)/ p_T(\ell_1)$ and $H_T({\rm jets})/M_{eff}$. }
\label{fig:opt-3l}
\end{figure}

Let us consider the $3\ell$ signal arising from the $WZ$ channel and the corresponding SM $WZ$ background\footnote{We do not employ a different and more dedicated strategy for the benchmarks contributing to the 3$\ell$ signature mainly through the $Wh$ channel.}. After requiring large values for MET, $M_{eff}^\prime$ and $M_T(W)$, the background events typically accompany a harder radiation (see left panel of \Fig{fig:opt-3l}). The peak of the jet energy fraction distribution, $H_T({\rm jets})/M_{eff}$,  is sharp and located at around 0.5 in the background, which means that all the remaining leptons and MET are recoiling against the radiated jets, in such a way that the total energy is balanced. On the other hand, the jet energy fraction is small for signal events, for which the leptons are already boosted thanks to the large mass splitting between NLSP and LSP. This is an interesting feature that generally appears in high-mass searches. Typically, the jet veto has been designed to suppress backgrounds with jets coming from the several particle decays, but now the jet veto can be very useful even to suppress backgrounds as $WZ$, in which jets only come from radiation. In our analysis we require the jet energy fraction, $H_T({\rm jets})/M_{eff}$, to be small to suppress the $WZ$ background.

In addition, the preferred configuration for the $WZ$ background generating high enough $M_T(W)$ is the leading lepton (or the single neutrino) particularly boosted by being aligned with the boost direction of the $W$. On the other hand, signal events contain several missing particles, neutrinos and neutralinos, thus $M_T(W)$ is not  upper limited by $M_W$ and can be easily large. As a result of the lower cut on $M_T(W)$, the background lepton $p_T$'s are more hierarchical as can be seen in the right panel of \Fig{fig:opt-3l}. 
In our analysis, we will impose a lower cut on the lepton $p_T$ ratio, $p_T(\ell_2)/p_T(\ell_1)$.

The lepton $p_T$ ratio and the variable $M_{eff}^\prime$ turn out to be more optimal for our analysis than the mere $M_{eff}$. This is partly because the information of $p_T(\ell_1)$ is used only by the lepton $p_T$ ratio but not by $M_{eff}^\prime$, which makes the two variables more independent and a better optimization possible. 
With these additional variables, the exclusion mass reach is improved by 400--800 GeV compared to the result based solely on $M_{eff}$, MET and $M_T(W)$.

\medskip

\begin{figure}[t] 
\includegraphics[width=0.95\textwidth]{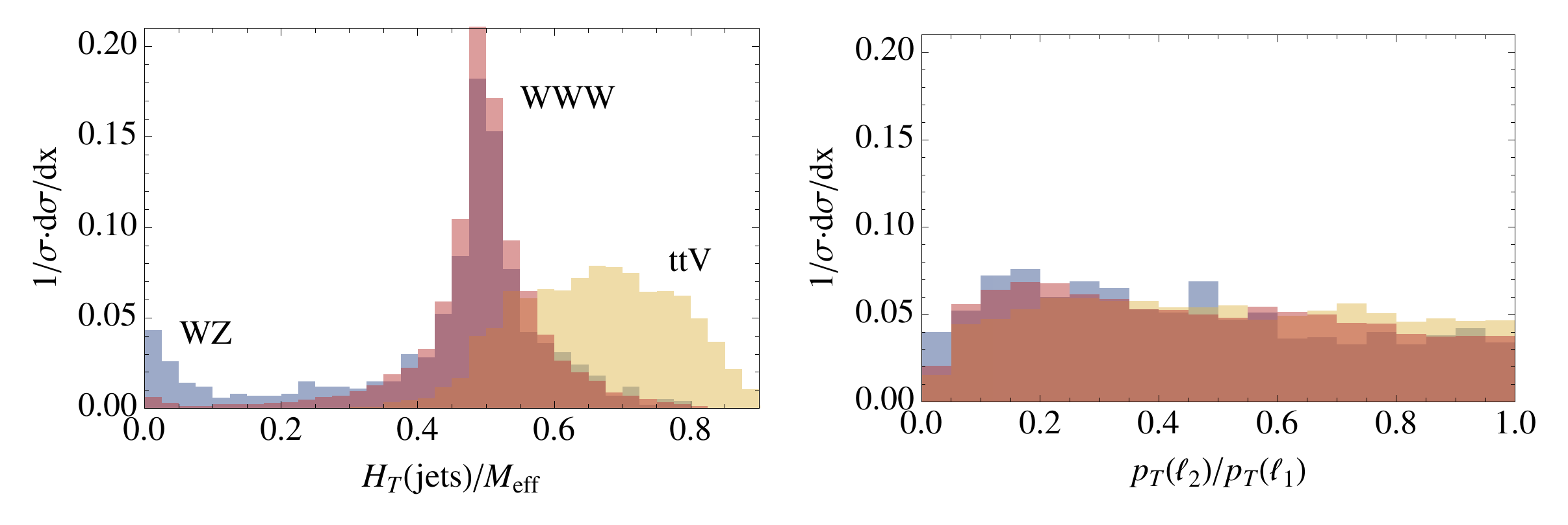}
\caption{Normalized distributions for the three background processes we consider in our analysis: dominant $WZ$ background (blue) and subdominant triboson $WWW$ (yellow) and $t\bar t V$ (red). The applied cuts are same as in \Fig{fig:opt-3l}.}
\label{fig:opt-3l-bkgd}
\end{figure}

These additional variables are also useful to make sure that additional backgrounds from triboson and $t\bar t V$ remain small.
The $t\bar t V$ 
background produces many jets from the top and vector boson decays, thus it has significant jet energy fraction as shown in the left panel of \Fig{fig:opt-3l-bkgd}. Therefore, the upper cut on the jet energy fraction can suppress efficiently the $t\bar t V$ backgrounds.
In contrast, the triboson background, $WWW$, does not produce any jets from the vector boson decays and needs hard radiation to reach a high $M_{eff}^\prime$, similarly to the $WZ$ background. Thus, they similarly have a sharp peak in the distribution of the jet energy fraction at 0.5 as shown in \Fig{fig:opt-3l-bkgd}. Nevertheless, $WWW$ has a sharper peak at 0.5 and a smaller tail at small jet energy fraction. The accumulation of $WZ$ events at small jet energy fraction is due to the $M_T(W)$ cut preferring a boosted leading lepton and/or neutrino. The $M_T(W)$ of  $WWW$, however, is not bounded from above by $M_T(W)$ and the lower cut on $M_T(W)$ does not induce the accumulation. Due to this difference, the upper cut on the jet energy fraction can suppress triboson backgrounds even more efficiently than the diboson background. 

Finally, in the right panel of \Fig{fig:opt-3l-bkgd}, we also check that the distributions for $p_T(\ell_2)/p_T(\ell_1)$ are very similar for all backgrounds. This demonstrates the goodness of a lower cut on $p_T(\ell_2)/p_T(\ell_1)$ suppressing all backgrounds. 

\medskip

\noindent
\underline{ OSDL Search: }

Let us consider the signal arising from the $WW$ channel and the dominant SM $W^+W^-$ background. 
First of all, our OSDL baseline cuts are defined with explicit jet vetos. Thus the jet energy fraction cut is not used in this analysis.

\begin{figure}[t] 
\includegraphics[width=0.95\textwidth]{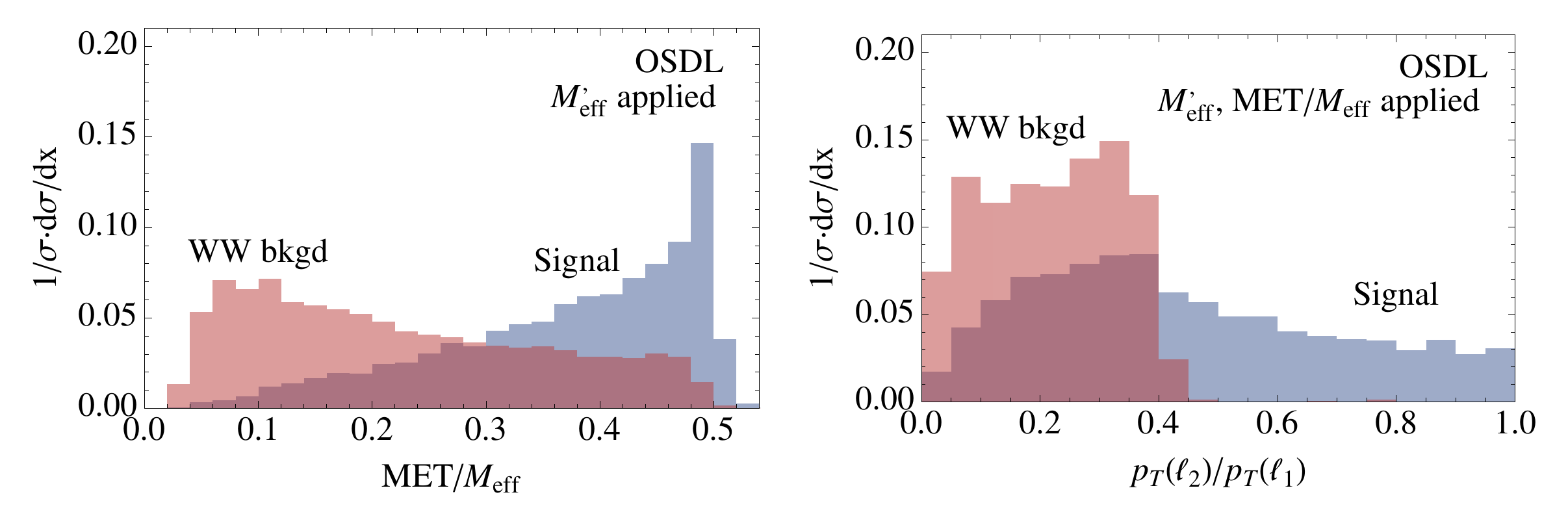}
\caption{Distributions for the  MET/$M_{eff}$ ratio {(left panel)}  and the lepton $p_T$ ratio {(right panel)}. In the left panel, baseline and $M_{eff}^\prime$ cuts are applied while MET/$M_{eff}$ is applied only in the right panel. Discovery cuts in Table~\ref{tab:WB-osdl-cutflow} are applied. $WW$ signal (blue) is from 2 TeV NLSP and massless LSP, and the background (red) is the SM $WW$ process. 
}
\label{fig:opt-osdl}
\end{figure}

We first require a large $M_{eff}^\prime$. 
The missing energy fraction, $\MET/M_{eff}$, is then a useful discriminator between signal and background. 
As signal events contain more missing particles, they tend to have a larger $\MET/M_{eff}$  (see left panel of \Fig{fig:opt-osdl}). Once a lower cut on the missing energy fraction is applied, a neutrino from the $WW$ background is likely aligned with the boost direction of its mother $W$ so that large MET is obtained. Consequently, the charged lepton from the same side stays soft, and the $p_T$ of the two charged leptons tend to be hierarchical as shown on the right panel of \Fig{fig:opt-osdl}.  As will be shown in Table~\ref{tab:WB-osdl-cutflow}, numerically optimized cuts on these ratio variables make the $W^+W^-$ and $WZ$ backgrounds similar in size. 

Meanwhile, transverse mass variables are often used in OSDL searches in analogy of the $M_T(W)$ in $3\ell$ searches. 
For example, the CMS $h \to WW^* \to 2\ell 2\nu$ analysis uses the transverse mass between MET and the lepton pair~\cite{Chatrchyan:2013iaa}. We find that using MET and this transverse mass is not any better in rejecting the $W^+W^-$ background. The jet veto implies a simple momentum conservation among the two charged leptons and the missing particles, $\vec{p}_T(\ell_1) + \vec{p}_T(\ell_2 ) \, \approx \, -\vec{\MET} $. Thus, $M_T(\MET,\ell \ell) \, \approx \, 2 \MET$: in this respect, the transverse mass is redundant. Nevertheless, the transverse mass is useful in rejecting non-negligible $WZ$ backgrounds.
We thus apply a lower cut on this transverse mass. Finally, we comment that the $t\bar{t}$ background has similar distributions of these variables as those of the $W^+W^-$ background because $t\bar{t}$ events are essentially reduced to $W^+ W^-$ events after jet veto. We have checked that the $t\bar t$ background is subdominant with our cuts.

With these additional variables, we can improve the mass reach by about 200-500 GeV compared to a simple MET and $M_{eff}$ analysis.

\medskip

\noindent
\underline{ SSDL Search: }

\begin{figure}[t] 
\includegraphics[width=0.95\textwidth]{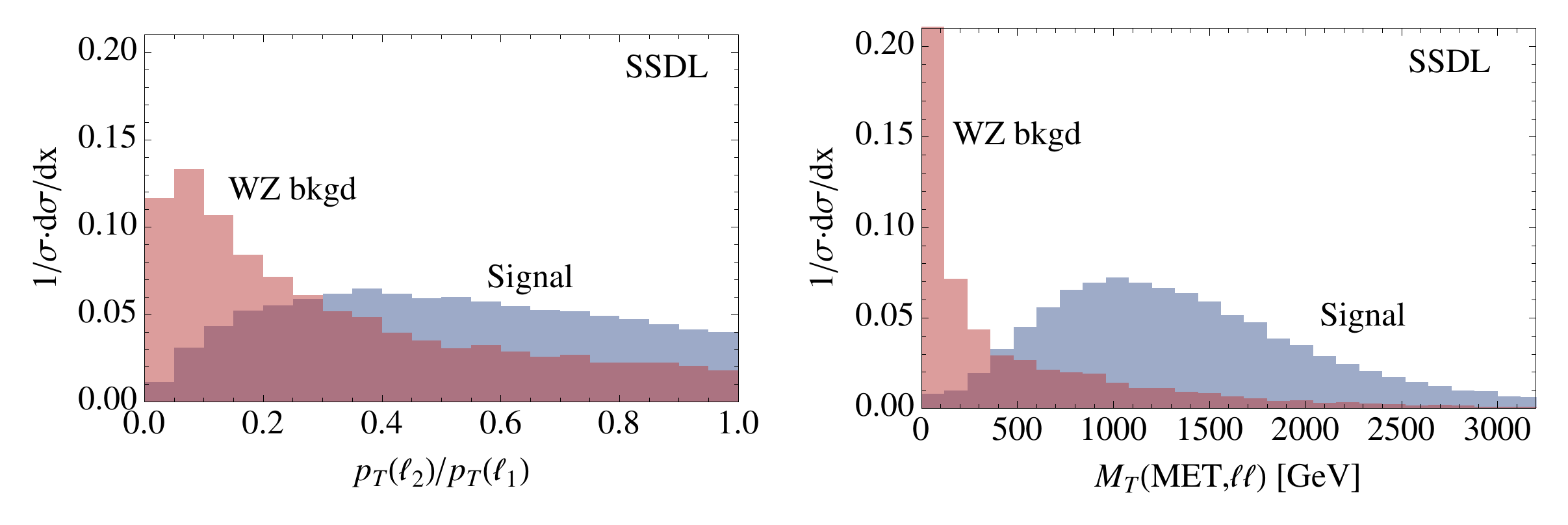}
\caption{Distributions for the lepton $p_T$ ratio {(left panel)} and $M_T(\MET,\ell \ell)$ {(right panel)}: the $W^\pm W^\pm$ signal (blue) is from 2 TeV NLSP and massless LSP, and the background (red) is the SM $WZ$ process. Discovery cuts listed in Table~\ref{tab:WH-ssdl-cutflow} except for the ones drawn here are applied.}
\label{fig:opt-ssdl}
\end{figure}

Let us compare the signal arising from the same-sign $W$ pair channel, $W^\pm W^\pm$, with the main SM background $WZ$.
The most important variable distinguishing between them is
the transverse mass variable, $M_T(\MET, \ell \ell)$. As shown in the right panel of \Fig{fig:opt-ssdl}, the background distribution is peaked at smaller values. The difference is more pronounced with heavier NLSPs.
As, in the background, the second lepton is hierarchically softer than the leading lepton (see left panel of \Fig{fig:opt-ssdl}), 
the transverse mass is approximately $M_T(\MET, \ell \ell) \, \approx \, M_T(\MET, \ell_1) \, \approx \, M_T(W)$, and thus has a strong drop at $M_W$. We apply a lower bound on this variable.

%%%
\subsection{Cut Optimization} \label{sec:opt-cut}

\begin{table}[t] \centering
\begin{tabular}{c| c| c c c| c c}
\hline \hline
Cuts & Signal & $WZ$ & $t\bar t V$ & $VVV$ & $S/B$ & $S/\sqrt{B}$ \\
\hline 
Baseline+mSFOS($Z$) & 18 & $7.3 \times 10^5$ & $6.7 \times 10^4 $ & $4.0 \times 10^3$ &  0.00002 & 0.02 \\
$M_{eff}^\prime > 2550$  & 11 & 8300 & 660 & 140 & 0.0013 & 0.12 \\
$\MET>1300, M_T(W)>300$ & 8.8 & 39 & 5.7 & 8.8 & 0.16 & 1.2 \\
$H_T({\rm jets})/M_{eff} < 0.25$ & 8.0 & 15 & 0.74 & 0.91 & 0.48 & 2.0 \\
$p_T(\ell_2) / p_T(\ell_1) > 0.25 $ & 7.4 & 5.8 & 0.59 & 0.71 & 1.0 & 2.8 \\
\hline\hline
\end{tabular}
\caption{Cut flow of the $3\ell$ search for a model with a 3 TeV Wino NLSP and massless Higgsino LSP. The $WZ$ channel is the dominant contribution to the signal. Number of events are calculated with 3/ab of data. Dimensionful cuts are in units of GeV. }
\label{tab:WH-3l-cutflow}\end{table}

\begin{table}[t] \centering
\begin{tabular}{c| c |c c c c| c c}
\hline \hline
Cuts & Signal & $WW$ & $WZ$ & $t\bar t V$ & $VVV$ & $S/B$ & $S/\sqrt{B}$ \\
\hline 
Baseline+mSFOS($Z$) & 140 &  $5.8 \times 10^5$ & 8400 & 460 & 610 & 0.00023 & 0.18\\ 
$M_{eff}^\prime > 1100$  & 61 & 620 & 55 & 0.049 & 2.8 & 0.090 & 2.3 \\
$M_T(\MET, \ell \ell)>1100$ & 55 & 210 & 22 & 0 & 1.5 & 0.23 & 3.6 \\
$\MET/M_{eff} > 0.36$ & 38 & 48 & 14 & 0 & 0.74 & 0.60 & 4.8 \\
$p_T(\ell_2) / p_T(\ell_1) > 0.24 $ & 27 & 10 & 7.6 & 0 & 0.35 & 1.5 & 6.4 \\
\hline\hline
\end{tabular}
\caption{Cut flow of the OSDL search for a model with a 2 TeV Wino NLSP and massless Bino LSP, Case 5 (see Tab.~\ref{tab:WB-cases} for the definition). The $WW$ channel is the dominant contribution to the signal. }
\label{tab:WB-osdl-cutflow}\end{table}

For each benchmark that we simulate, we optimize the cuts on the variables discussed in the previous subsections to maximize the statistical significance, $S/\sqrt{B}$, where the $S(B)$ are the number of signal (background) events with the assumed luminosity (3000 fb$^{-1}$). To this end, ${\cal O}(10)^n$-size arrays of cut efficiencies are generated for each process, where the $n$ is the number of variables that we optimize. We always require at least 5 signal events after all cuts. For the $3\ell$ and OSDL searches, we do not assume any systematic uncertainties and do not vary background normalization. For the SSDL search, instead, reducible backgrounds can be especially large and difficult to estimate. According to the current SSDL searches with jet vetos~\cite{Khachatryan:2014qwa}, the reducible backgrounds may be of similar size as the ones from diboson production. Thus, to take them into account, we conservatively multiply the simulated backgrounds by 2 in the SSDL analysis -- this is denoted by $N=2$ in the notation of \Eq{eq:errordef}-- and $S/\sqrt{NB}$ is maximized in our optimization. 

In Tables~\ref{tab:WH-3l-cutflow}--\ref{tab:WB-3lh-cutflow}, we present our optimal cuts and results for the all multilepton searches using benchmark scenarios that will be discussed in \Sec{sec:prospects}\footnote{Here and in the following tables, we present benchmarks with massless LSP. Results would basically not change choosing $m_{\rm{LSP}}\sim 100$ GeV, since we are still in a regime of $m_{\rm{LSP}}\ll m_{\rm{NLSP}}$.}. After all cuts, diboson backgrounds are generally dominant, while $t\bar t V$ and tribosons can only be important for $Wh$-dominated $3\ell$ searches, as shown in Table~\ref{tab:WB-3lh-cutflow}. As discussed, we checked that the SM $t\bar{t}$ background is smaller than $WW$ and $WZ$ backgrounds in the OSDL search in Table~\ref{tab:WB-osdl-cutflow}. We also checked that the irreducible SM $Wh$ background is small giving 6 events after all cuts in Tab.~\ref{tab:WB-3lh-cutflow}.
In particular, in the last one or two steps of each cut flow table, we also show the results of applying cuts on the new ratio variables introduced in the previous subsection. 

In the next section, we will present discovery and exclusion prospects based on the strategies and cuts described in this section.

\begin{table}[t] \centering
\begin{tabular}{c| c | c c c| c c}
\hline \hline
Cuts & Signal & $WZ$ & $t\bar t V$ & $VVV$ & $S/NB$ & $S/\sqrt{NB}$ \\
\hline 
Baseline+mSFOS($Z$) & 115 & $ 1.8 \times 10^5$ & $ 2.3 \times 10^4 $ & 1900 & 0.00028 & 0.18 \\
$M_{eff}^\prime > 980$  & 80 & $2.1 \times 10^4$ & 5200 & 420 & 0.0015 & 0.35 \\
$\MET > 660$ & 59 & 2100 & 240 & 87 & 0.012 & 0.84 \\
$M_T(\MET, \ell \ell)>1520$ & 40 & 130 & 13 & 5.6 & 0.14 & 2.4 \\
$p_T(\ell_2) / p_T(\ell_1) > 0.2 $ & 32 & 56 & 2.8 & 2.5 & 0.26 & 2.9 \\
\hline\hline
\end{tabular}
\caption{Cut flow of the SSDL search for the model with a 2 TeV Wino NLSP and massless Higgsino LSP. The background normalization, $N=2$ in \Eq{eq:errordef}, is chosen to account for reducible backgrounds from fakes and mis-identifications. The $WW$ channel is the dominant contribution to the signal. }
\label{tab:WH-ssdl-cutflow}\end{table}

\begin{table}[t] \centering
\begin{tabular}{c| c | c c c| c c}
\hline \hline
Cuts & Signal & $WZ$ & $t\bar t V$ & $VVV$ & $S/B$ & $S/\sqrt{B}$ \\
\hline 
Baseline+mSFOS($Z$) & 140 &  $4.4 \times 10^4$ & $2.9 \times 10^4$ & 4700 & 0.0018 & 0.50\\
$M_T(W) > 100$ & 130 & 6500 & $1.6 \times 10^4$ & 2700 & 0.0051 & 0.80 \\
$M_{eff}^\prime>600, \MET>400$ & 56 & 340 & 530 & 480 & 0.042 & 1.5 \\
$H_T({\rm jets})/M_{eff} < 0.25$ & 40 & 148 & 70 & 93 & 0.13 & 2.3 \\
\hline\hline
\end{tabular}
\caption{Cut flow of the $3\ell$ search for the model with a 1 TeV Wino NLSP and massless Bino LSP, Case 3. The $Wh$ channel is the dominant contribution to the signal.}
\label{tab:WB-3lh-cutflow}\end{table}

%%%%%%%
\section{Prospects of a 100 TeV Collider}\label{sec:prospects}

We present results for the following cases:
\begin{itemize}
\item Wino NLSP and Higgsino LSP (Wino-Higgsino) : $M_1\gg M_2>\mu$,
\item Higgsino NLSP and Wino LSP (Higgsino-Wino) : $M_1\gg \mu>M_2$,
\item Higgsino NLSP and Bino LSP (Higgsino-Bino) : $M_2\gg \mu>M_1$,
\item Wino NLSP and Bino LSP (Wino-Bino) : $\mu\gg M_2>M_1$.
\end{itemize}
The heaviest electroweakino mass is always fixed to 5 TeV.
We do not study the cases of (mainly) Bino NLSPs because of their small production rates. 
We do not use simplified models to present results. We rather take into account all the relevant branching ratios to gauge bosons and Higgs bosons. Only in the cases of light Higgsinos (NLSPs or LSPs), the dependence of the results on additional parameters ($t_\beta$ and signs of gaugino and Higgsino masses) vanishes. This is because the Higgsino system consists of two nearly degenerate indistinguishable neutralinos, $\tilde H_u^0$ and $\tilde H_d^0$, and summing their contributions removes such dependences. Consequently, we always have BR($N_{\rm NLSP} \to N_{\rm LSP} Z) =  {\rm BR}(N_{\rm NLSP} \to N_{\rm LSP} h)$ in the split parameter space~\cite{Jung:2014bda}\footnote{There are interesting exceptions from models with weakly interacting LSPs such as axinos or gravitinos due to slow decays of heavier Higgsinos~\cite{Barenboim:2014kka}.}. Specifically, we have BR($N_{\rm NLSP} \to N_{\rm LSP} Z) =  {\rm BR}(N_{\rm NLSP} \to N_{\rm LSP} h) \approx 1/4, 1/2, 1/3$ for the Wino-Higgsino, Higgsino-Bino and Higgsino-Wino scenarios, respectively. For the other case of Wino-Bino, results depend sensitively on the additional parameters (see Sec.~\ref{sec:WB}).

\subsection{Higgsino LSP} \label{sec:WH}

When the Higgsino is the LSP, the production of Wino NLSPs can be used to probe the model. Multi-lepton signals arise from the following processes:
\bei
\item The $3\ell$ arises mainly from $N_3 C_2 \to WZ + N_{1,2} N_{1,2}, \, WZ + C_1 C_1$ and $ C_2 C_2 \to WZ + N_{1,2} C_1$.
\item The OSDL arises mainly from $C_2 C_2 \to W^+ W^- + N_{1,2} N_{1,2}$ and $N_3 C_2 \to W^+ W^- + N_{1,2} C_1$.
\item The SSDL arises mainly from $N_3 C_2 \to W^\pm W^\pm + N_{1,2} C_1$.
\eei

\begin{table}[t] \centering
\begin{tabular}{c| c| c c c}
\hline \hline
\multicolumn{5}{c}{ Wino-Higgsino. $\sigma(N_3 C_2)=120$fb,  $\sigma(C_2 C_2)= 59.4$fb , $\sigma(N_3 N_3) \simeq 0$}\\
\hline \hline
intermediate dibosons & $\sigma$ (fb) & 3$\ell$ (ab) & OSDL (ab) & SSDL (ab) \\
\hline 
$WZ$ & 46 fb   & 124  & 5.3 & 52.8 \\
$Wh$ & 44 fb   & 0.6  & 0.7 & 3.6 \\
$W^+ W^-$ & 31 fb   & -- & 48.5 & -- \\
$W^\pm W^\pm$ & 16 fb  & -- & -- & 394 \\
$ZZ$ & 11 fb & 6.6 & 0.1 & 0.5 \\
\hline\hline
\end{tabular}
\caption{Multi-lepton signal rates are decomposed into each diboson channel contribution. The second column shows the total (intermediate) diboson rates produced from all possible NLSP pair decays. In the last three columns, we show cross sections of each diboson channel in multilepton final states after all discovery cuts are applied; cuts are listed in App.~\ref{sec:cutstable}. The chosen benchmark is 1 TeV Wino NLSP and massless Higgsino LSP. ``--'' indicates a contribution smaller than 0.05 ab. We do not show the 
$Zh$ and $hh$ channels since they are always small. }
\label{tab:WH-cont}\end{table}

In Table~\ref{tab:WH-cont}, we decompose the multi-lepton signal rates into each diboson channel contribution for a benchmark with a 1 TeV Wino NLSP and a massless Higgsino LSP. As mentioned, the $3\ell$, OSDL and SSDL channels get dominant contributions from the $WZ,\, W^+W^-$ and $W^\pm W^\pm$ diboson channels, respectively. In spite of the fact that ${\rm BR}(N_{\rm NLSP} \to N_{\rm LSP} h) \sim 0.25$, the $Wh$ channel contributions are subdominant in all final states because the Higgs's leptonic branching ratio, $h \to WW^* (ZZ^*) \to \ell \nu \ell \nu$, is small. Their contribution to the discovery reach is subdominant.

\begin{figure}[t] 
\includegraphics[width=0.49\textwidth]{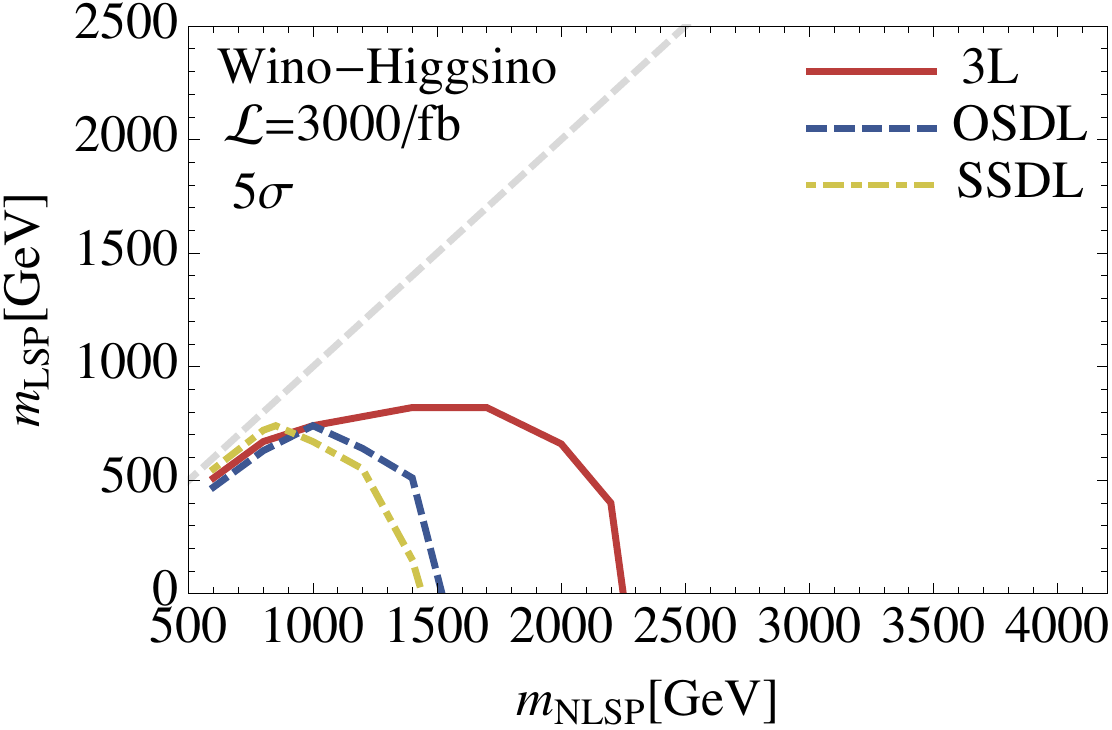}
\includegraphics[width=0.49\textwidth]{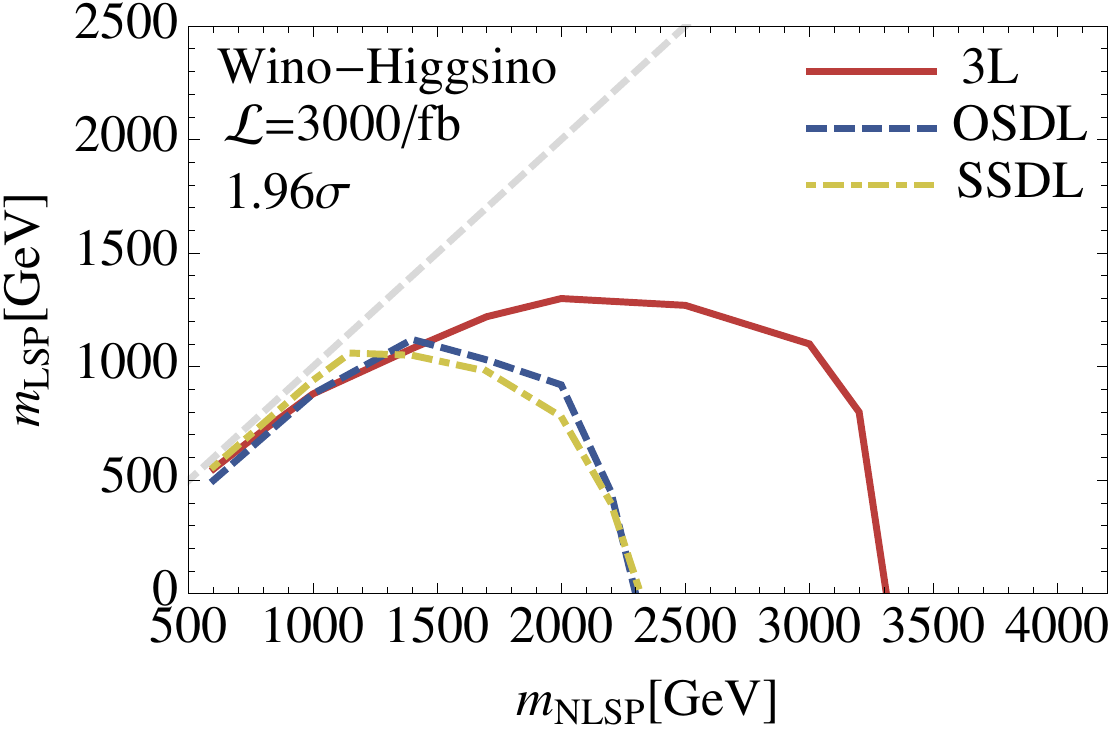}
\caption{$5\sigma$ discovery reaches {(left panel)} and 1.96$\sigma$ CL exclusion limits {(right panel)} of the Wino-NLSP and Higgsino-LSP model from the $3\ell$ (red solid), OSDL (blue dashed) and SSDL (green dot-dashed) searches. 
}
\label{fig:WH-reach}
\end{figure}

The corresponding reach is presented in \Fig{fig:WH-reach}. We do not specify our choice of additional parameters ($t_\beta$ and the sign of gaugino and Higgsino masses), since the branching ratios of the NLSP are model independent in this Higgsino LSP case. 
As expected, the $3\ell$ signature can probe the highest NLSP mass while the SSDL signature can be useful in the region with a smaller mass difference between the NLSP and the LSP.  

It is important to note that a 100 TeV collider with $3000/$fb data will be able to exclude Higgsino dark matter ($m_{\rm{LSP}}\sim 1 $ TeV) for 
Winos lighter than about 3.2 TeV and not too close in mass to the Higgsino. Achieving the significance needed for discovery of a 1 TeV Higgsino, however, is expected to be rather difficult (see left panel of \Fig{fig:WH-reach}). Ref.~\cite{Low:2014cba} shows that monojet and disappearing charged track searches at a 100 TeV collider also can have difficulties in probing 1 TeV Higgsino dark matter. In addition, Higgsino dark matter is a very challenging scenario to discover from the astrophysical side, since current astrophysical photon line/continuum searches lack  sensitivity to 1 TeV Higgsinos as well~\cite{Fan:2013faa}.

%%%
\subsection{Wino LSP} \label{sec:HW}

When the Wino is the LSP, Higgsino NLSPs can be used to probe the model. Multi-lepton signals arise from the following processes:
\bei
\item The $3\ell$ arises mainly from $N_{2,3} C_2 \to WZ + N_1 N_1, \, WZ + C_1 C_1$ and $ C_2 C_2 \to WZ + N_1 C_1$ and $N_2 N_3 \to WZ + N_1 C_1$.
\item The OSDL arises mainly from $C_2 C_2 \to W^+ W^- + N_1 N_1$ and $N_{2,3} C_2 \to W^+ W^- + N_1 C_1$ and $N_2 N_3 \to W^+ W^- N_1 C_1$.
\item The SSDL arises mainly from $N_{2,3} C_2 \to W^\pm W^\pm + N_1 C_1$ and $N_2 N_3 \to W^\pm W^\pm C_1 C_1$.
\eei
In Table~\ref{tab:HW-cont}, we decompose multi-lepton signal rates into each diboson contribution. The qualitative discussion in this table is the same as for Table~\ref{tab:WH-cont}.

\begin{table}[t] \centering
\begin{tabular}{c| c| c c c}
\hline \hline
\multicolumn{5}{c}{ Higgsino-Wino. $\sigma(N_{2,3} C_2)=60.0$fb,  $\sigma(C_2 C_2)= 17.3$fb, $\sigma(N_2 N_3)=16.0$fb  }\\
\hline \hline
intermediate dibosons & $\sigma$ (fb) & 3$\ell$ (ab) & OSDL (ab) & SSDL (ab) \\
\hline 
$WZ$ & 24 fb   & 65.3  & 2.8 & 27.8 \\
$Wh$ & 24 fb   & 0.3  & 0.4 & 1.8 \\
$W^+ W^-$ & 12 fb   & -- & 17.0 & -- \\
$W^\pm W^\pm$ & 10 fb  & -- & -- & 222 \\
$ZZ$ & 5.8 fb & 3.9 & 0.1 & 0.8 \\
\hline\hline
\end{tabular}
\caption{Same as in Table~\ref{tab:WH-cont} but for the 1 TeV Higgsino NLSP and massless Wino LSP benchmark scenario.}
\label{tab:HW-cont}\end{table}

\begin{figure}[t] 
\includegraphics[width=0.49\textwidth]{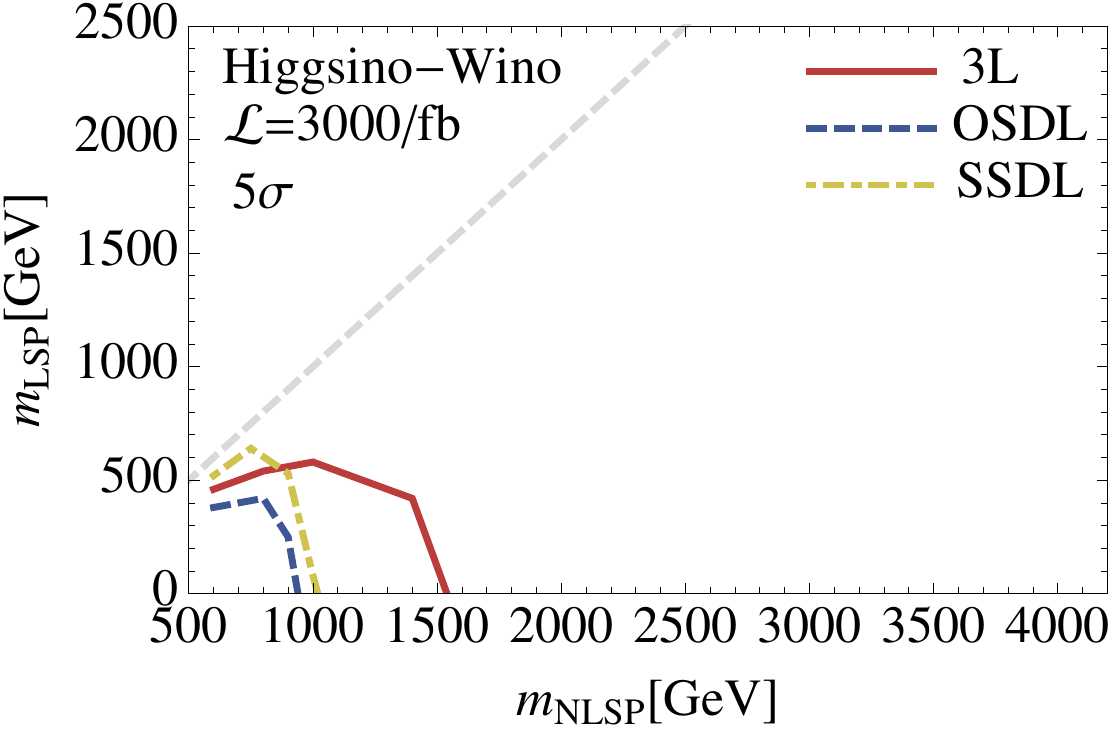}
\includegraphics[width=0.49\textwidth]{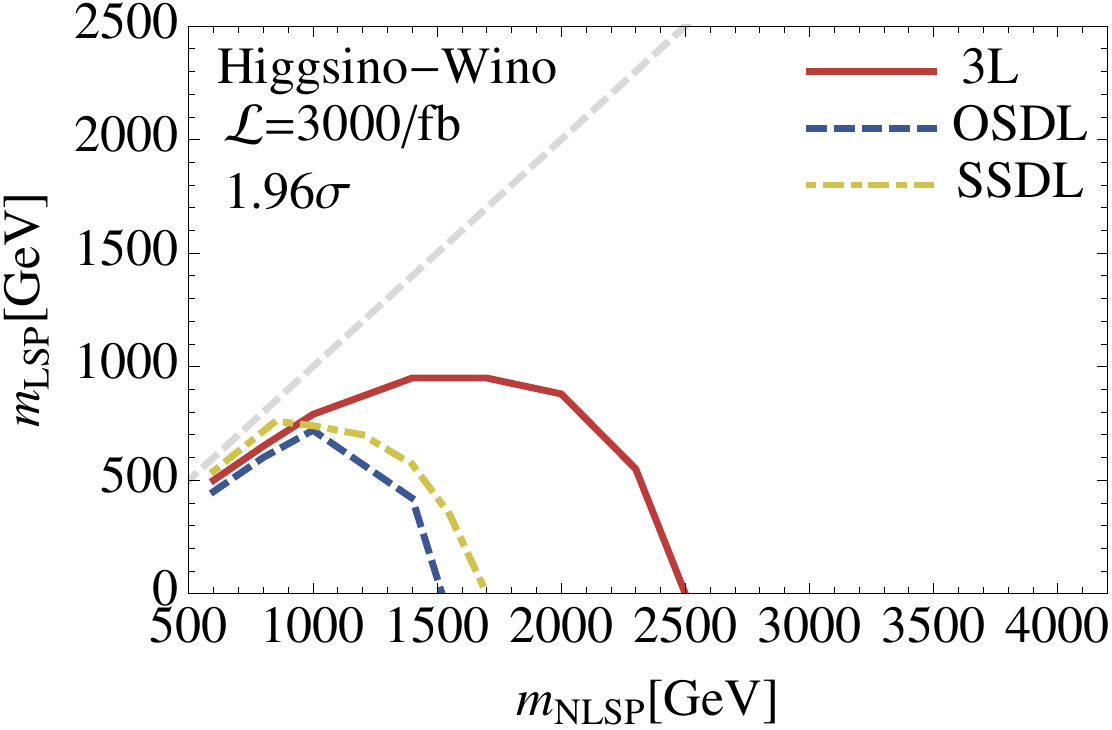}
\caption{Same as in \Fig{fig:WH-reach} but for the Higgsino-NLSP and Wino-LSP scenario. }
\label{fig:HW-reach}
\end{figure}

The reach is presented in \Fig{fig:HW-reach}. As we already discussed, the branching ratios of the NLSP pairs to $WZ$, $W^+ W^-$, $W^\pm W^\pm$ and $Wh$ are again independent of the choice of parameters, as Higgsinos are involved in the decay. The $3\ell$ signature can probe the highest NLSP mass, while the SSDL signature can be useful in the region with smaller mass difference. Compared to the Wino NLSP results shown in \Fig{fig:WH-reach}, the reach here is worse, mainly because Higgsino NLSP production cross sections are smaller than the Wino ones.

From the figure, we note that the multi-lepton NLSP searches cannot rule out or discern  $\sim 3$ TeV Wino thermal dark matter. 
Wino dark matter, however, is expected to be probed by monojet and disappearing charged track searches at 100 TeV~\cite{Low:2014cba}, as well as by astrophysical photon line/continuum searches~\cite{Cohen:2013ama,Fan:2013faa}.

%%%
\subsection{Bino LSP with Higgsino NLSP} \label{sec:HB}

When the Bino is the LSP, either Higgsino or Wino NLSPs can be used to probe the model. In this subsection, we first consider the Higgsino NLSP case since it is simpler to discuss. Multi-lepton signals arise from the following processes:

\begin{itemize}
\item The $3\ell$ arises mainly from $N_{2,3} C_1 \to WZ + N_1 N_1$.
\item The OSDL arises mainly from $C_1 C_1 \to W^+ W^- + N_1 N_1$.
\item The SSDL arises mainly from the $WZ$ channel by accidentally loosing one lepton: 
$N_{2,3} C_1 \to WZ + N_1 N_1$. The $W^\pm W^\pm$ channel is not produced as shown in Table~\ref{tab:HB-cont}.
\eei
Multi-lepton signal rates are decomposed into each diboson contribution in Table~\ref{tab:HB-cont}.

\begin{table}[t] \centering
\begin{tabular}{c| c| c c c}
\hline \hline
\multicolumn{5}{c}{ Higgsino-Bino. $\sigma(N_{2,3} C_2)=60.0$fb,  $\sigma(C_1 C_1)= 17.3$fb, $\sigma(N_2 N_3)=16.0$fb  }\\
\hline \hline
intermediate dibosons & $\sigma$ (fb) & 3$\ell$ (ab) & OSDL (ab) & SSDL (ab) \\
\hline 
$WZ$ & 30 fb   & 81.5  & 3.5 & 28.3 \\
$Wh$ & 30 fb   & 0.4  & 0.5 & 1.5 \\
$W^+ W^-$ & 17 fb   & -- & 27.3 & -- \\
$W^\pm W^\pm$ & 0 fb  & -- & -- & -- \\
$ZZ$ & 4.1 fb & 2.4 & 0.1 & 0.4 \\
\hline\hline
\end{tabular}
\caption{Same as in Table~\ref{tab:WH-cont} but for the 1 TeV Higgsino NLSP and massless Bino LSP benchmark scenario.}
\label{tab:HB-cont}\end{table}

The reach is presented in \Fig{fig:HB-reach} and is independent of the particular choice of additional parameters. The $3\ell$ channel is by far the best. The SSDL signature gives now a much weaker bound than the OSDL signature because SSDL arises only from $WZ$ by accidentally loosing one lepton. 
Higgsino NLSPs up to about 3 TeV and Bino LSPs up to about 1 TeV can be excluded at optimal points in the parameter space.

\begin{figure}[t] 
\includegraphics[width=0.49\textwidth]{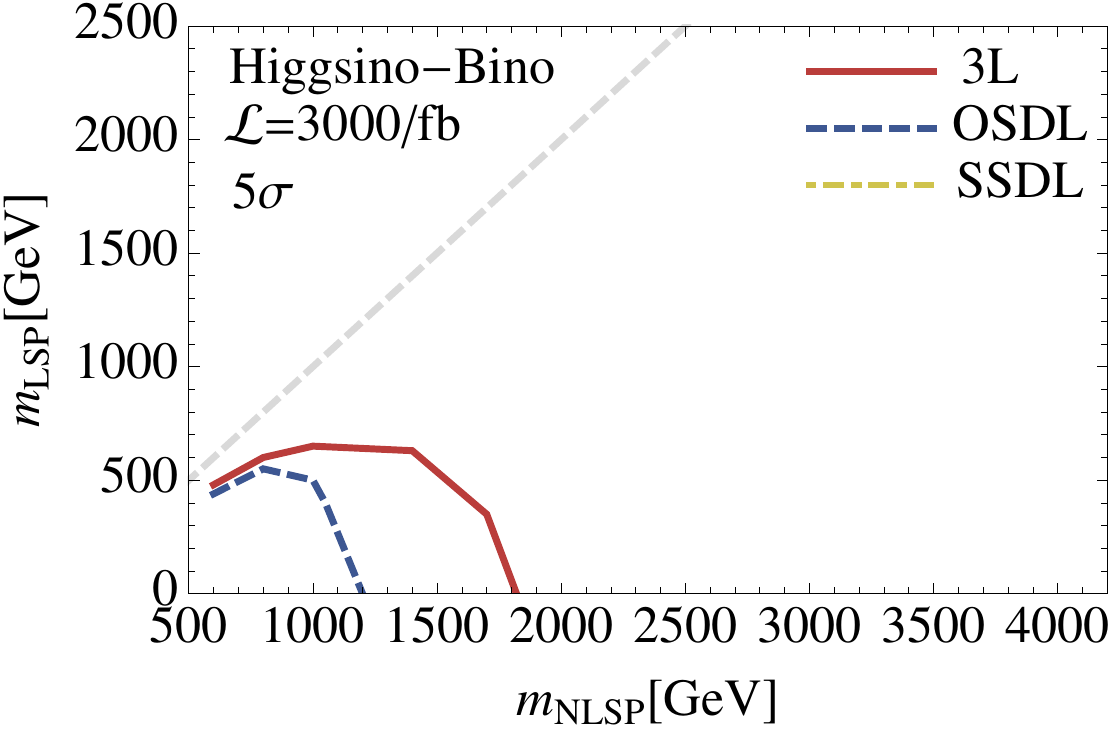}
\includegraphics[width=0.49\textwidth]{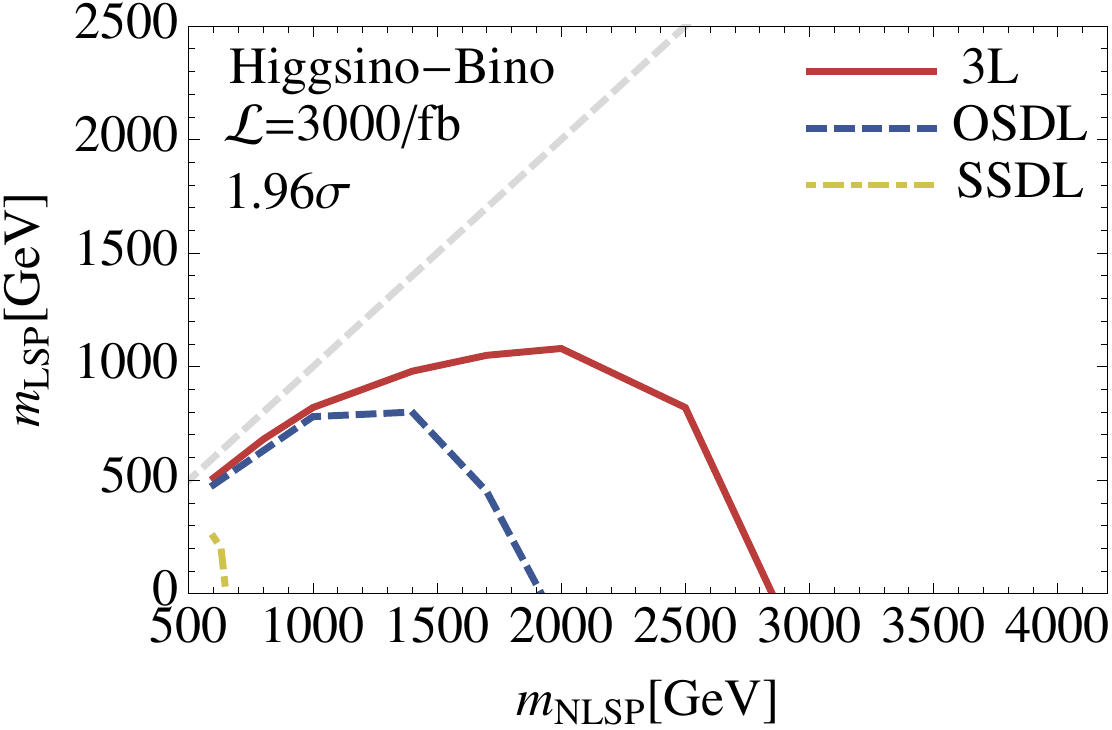}
\caption{Same as in \Fig{fig:WH-reach} but for the Higgsino-NLSP and Bino-LSP scenario.}
\label{fig:HB-reach}
\end{figure}

%%%%%%%%%%%%%%%%%%%%%%%%%%%
\subsection{Bino LSP with Wino NLSP} \label{sec:WB}

Wino NLSPs can also be used to probe the Bino LSP scenario. 
The multi-lepton signals arise from the following processes:
\bei
\item The $3\ell$ arises mainly from $N_2 C_1 \to WZ + N_1 N_1$.
\item The OSDL arises mainly from $C_1 C_1 \to W^+ W^- + N_1 N_1$.
\item The SSDL arises mainly from the $WZ$ channel by accidentally loosing one lepton: $N_2 C_1 \to WZ + N_1 N_1$. The $W^\pm W^\pm$ channel is not produced as shown in Table~\ref{tab:WB-cont}.
\eei

The branching ratios of Winos depend now sensitively on the choice of parameters:\footnote{Only two of the sign($M_1 M_2^*$), sign($ \mu M_2^*$) and sign($M_1 \mu^*$) are physical. The former two are most convenient choices to understand our numerical results. Without loss of generality, we assume that mass parameters are real and $M_1 \geq 0$.
} 
\beq
t_\beta, \,\, {\rm sign}(\mu M_2), \,\, {\rm and} \,\,\,  {\rm sign}(M_2 M_1).
\eeq
In this section, we fix $|\mu|=5$ TeV. 
In \Fig{fig:WB-3l-reach}, we collect our results for the $3\ell$ channel using the 
six sets of parameters listed in Table~\ref{tab:WB-cases}. The remaining two possible choices are not much qualitatively different from these choices.
\begin{table}[t] \centering 
\setlength{\tabcolsep}{7pt} \renewcommand{\arraystretch}{1.4}
\begin{tabular}{c  c || c c }
\hline \hline
\multicolumn{4}{c}{ $t_\beta, \, {\rm sign}(\mu M_2), \, {\rm sign}(M_2 M_1)$ }\\
\hline \hline
Case 1 : & $50, \, +, \, +$  &  Case 2 : & $50, \, -, \, +$ \\
\hline
Case 3 :  & $3, \, +, \, +$  &  Case 4 : & $3, \, -, \, +$ \\
\hline
Case 5 :  & $50, \, -, \, -$  &  Case 6 : & $3, \, -, \, -$ \\
\hline \hline
\end{tabular}
\caption{Benchmarks for the Wino-Bino model. The $\mu$ parameter is fixed to $|\mu|=5$ TeV, $M_1 \geq $0.}
\label{tab:WB-cases}\end{table}

\begin{table}[t] \centering
\begin{tabular}{c| c| c c c}
\hline \hline
\multicolumn{5}{c}{ Wino-Bino. $\sigma(N_2 C_1)=120$fb,  $\sigma(C_1 C_1)= 59.4$fb, $\sigma(N_2 N_2) \simeq 0$ }\\
\hline \hline
intermediate dibosons & $\sigma$ (fb) & 3$\ell$ (ab) & OSDL (ab) & SSDL (ab) \\
\hline 
$WZ$ & 88 fb   & 240  & 10.6 & 86.9 \\
$Wh$ & 32 fb   & 0.4  & 0.5 & 1.4 \\
$W^+ W^-$ & 59 fb   & -- & 93.9 & -- \\
$W^\pm W^\pm$ & 0 fb  & -- & -- & -- \\
$ZZ$ & 0 fb & -- & -- & -- \\
\hline\hline
\end{tabular}
\caption{Same as in Table~\ref{tab:WH-cont} but for the 1 TeV Wino NLSP and massless Bino LSP, Case 5 benchmark scenario.}
\label{tab:WB-cont}\end{table}

The OSDL and SSDL results are presented in \Fig{fig:WB-2l-reach}. Here we consider only one benchmark (Case 5) since the reach of the OSDL channel is rather model independent and the reach of the SSDL channel is weak. The OSDL channel receives the main contribution from chargino pair production and chargino pairs always lead to the $W^+W^-$ channel. It has the highest reach in this Wino-Bino model among all models we have investigated (note that this model has the highest rate for $W^+W^-$, see Table~\ref{tab:WB-cont}).

The OSDL can exclude up to about 3 TeV NLSPs. The $3\ell$ can exclude higher or lower masses depending on the parameters. At best, it can exclude $\sim$4.3 TeV NLSPs (Case 4 and 6) while only $\sim$1.3 TeV at worst (Case 3). 

\begin{figure}[t] 
\includegraphics[width=0.49\textwidth]{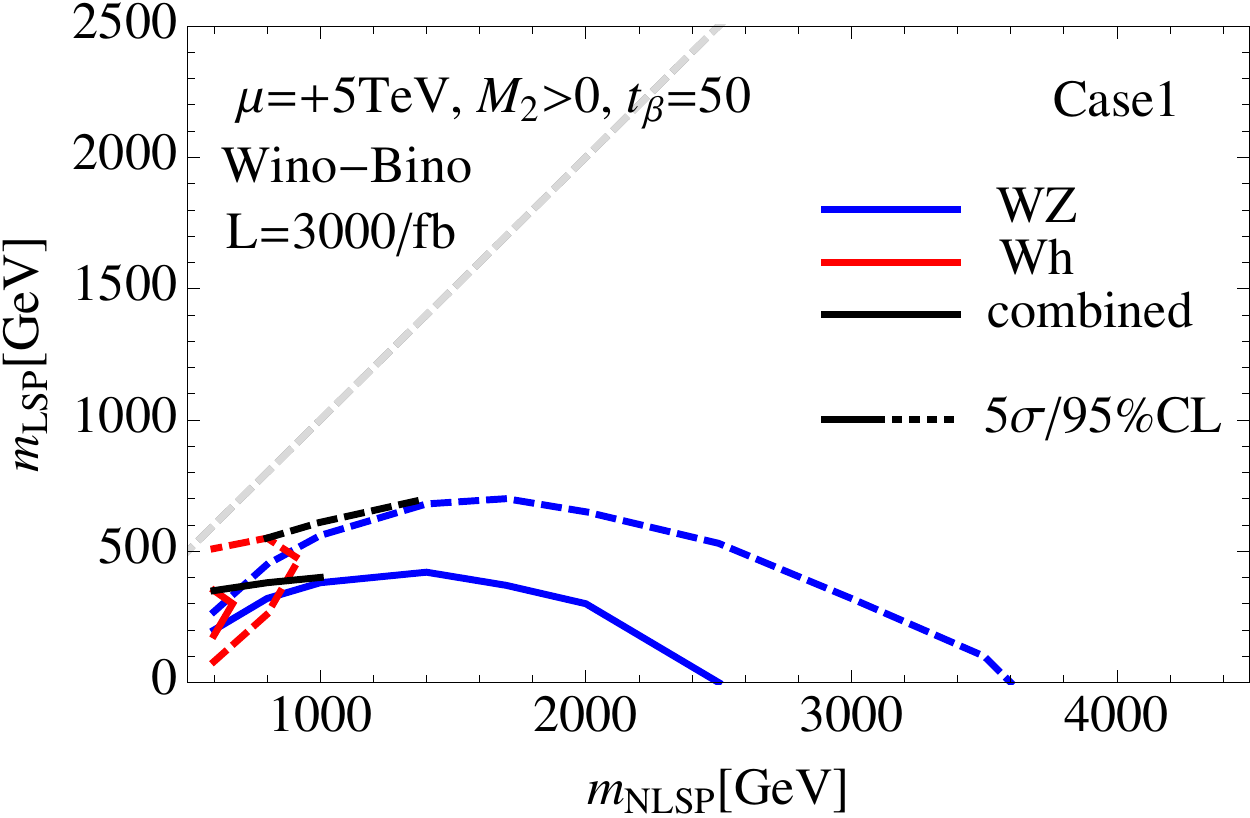}
\includegraphics[width=0.49\textwidth]{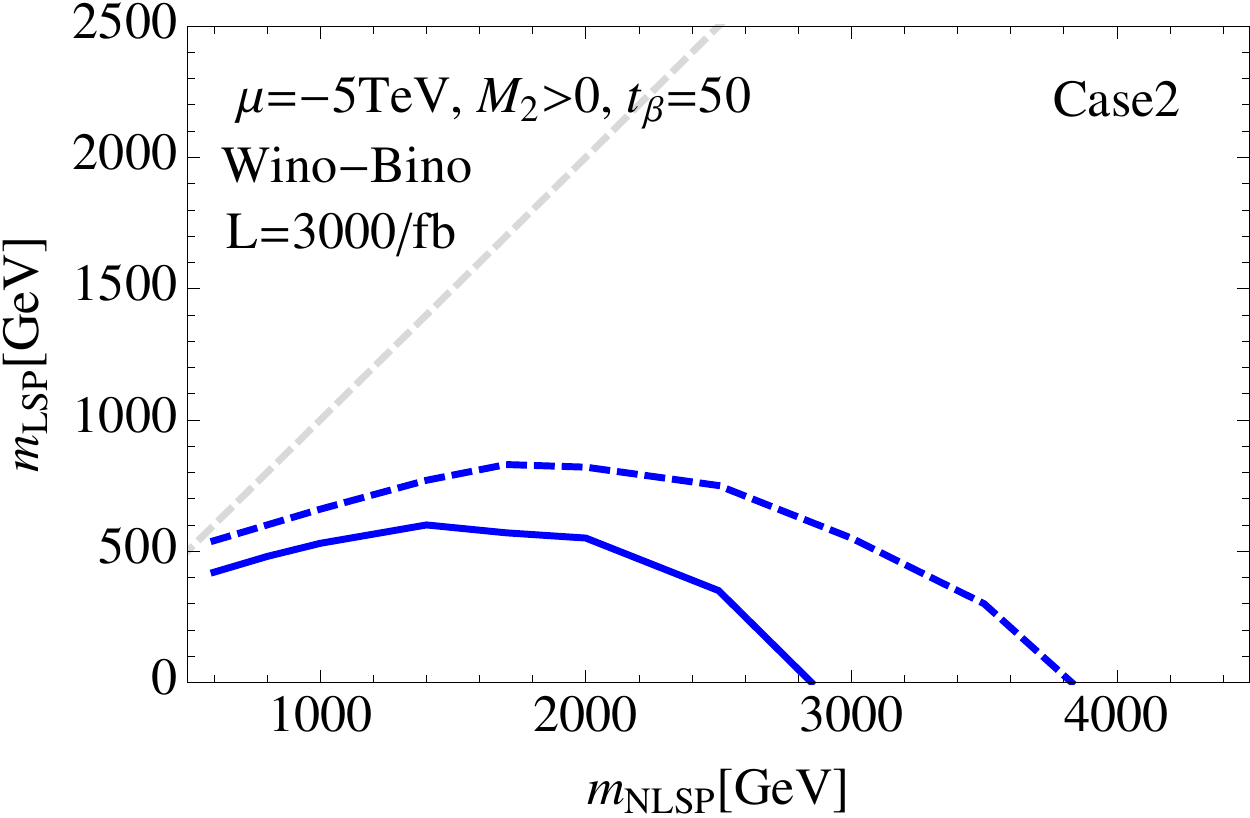}\\
\includegraphics[width=0.49\textwidth]{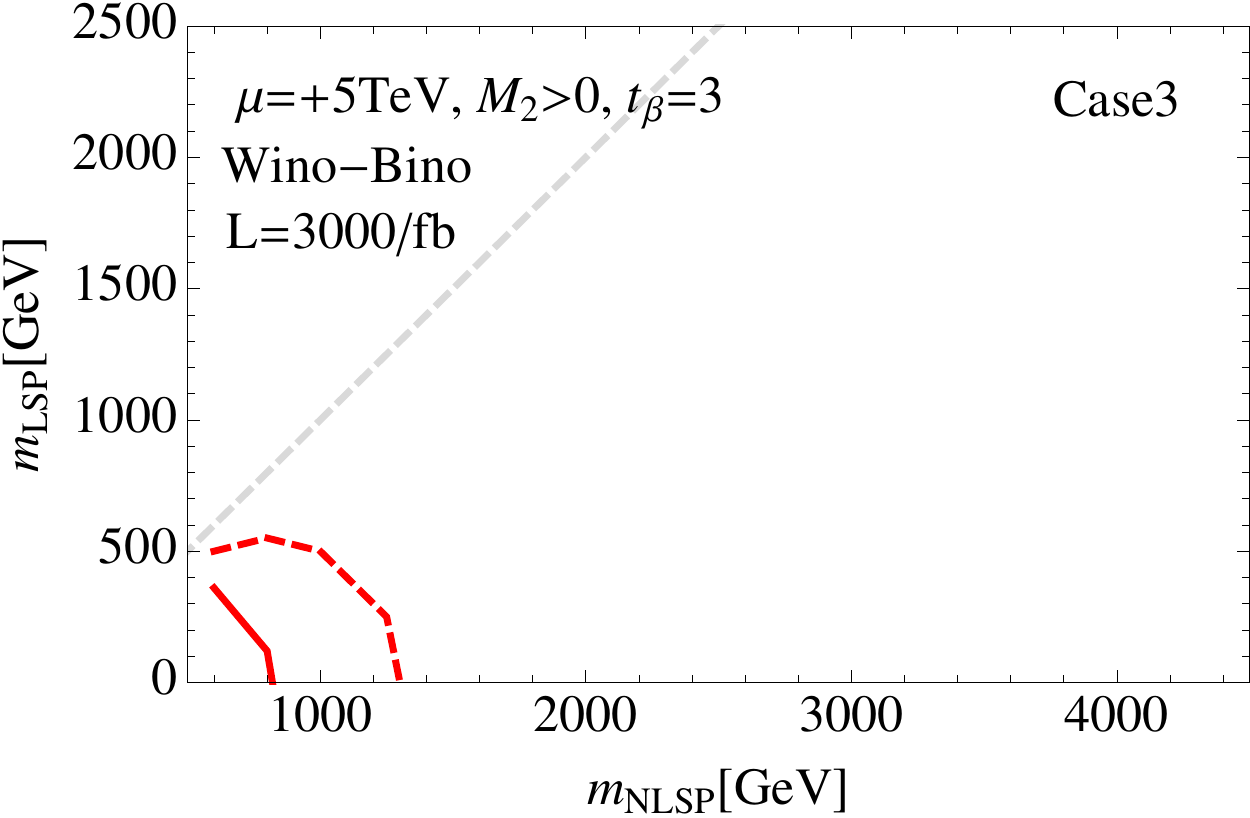}
\includegraphics[width=0.49\textwidth]{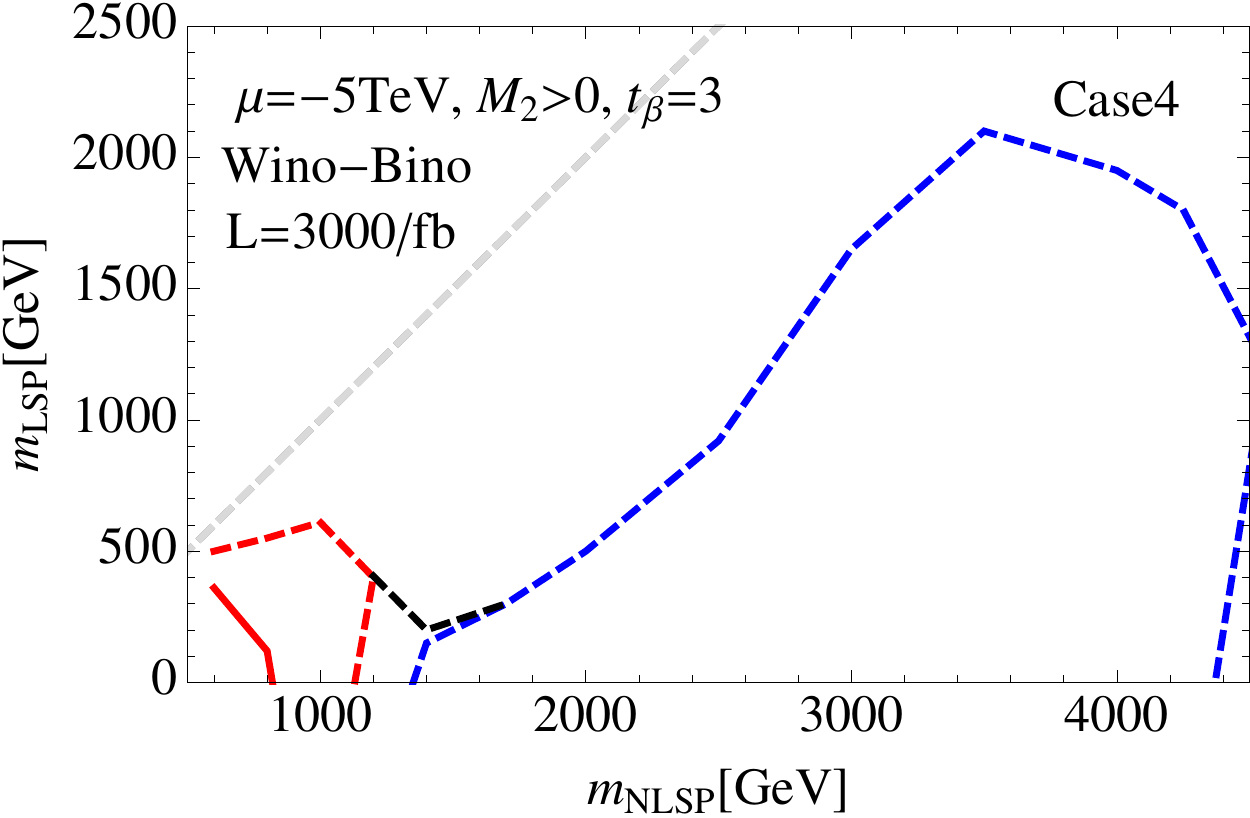}\\
\includegraphics[width=0.49\textwidth]{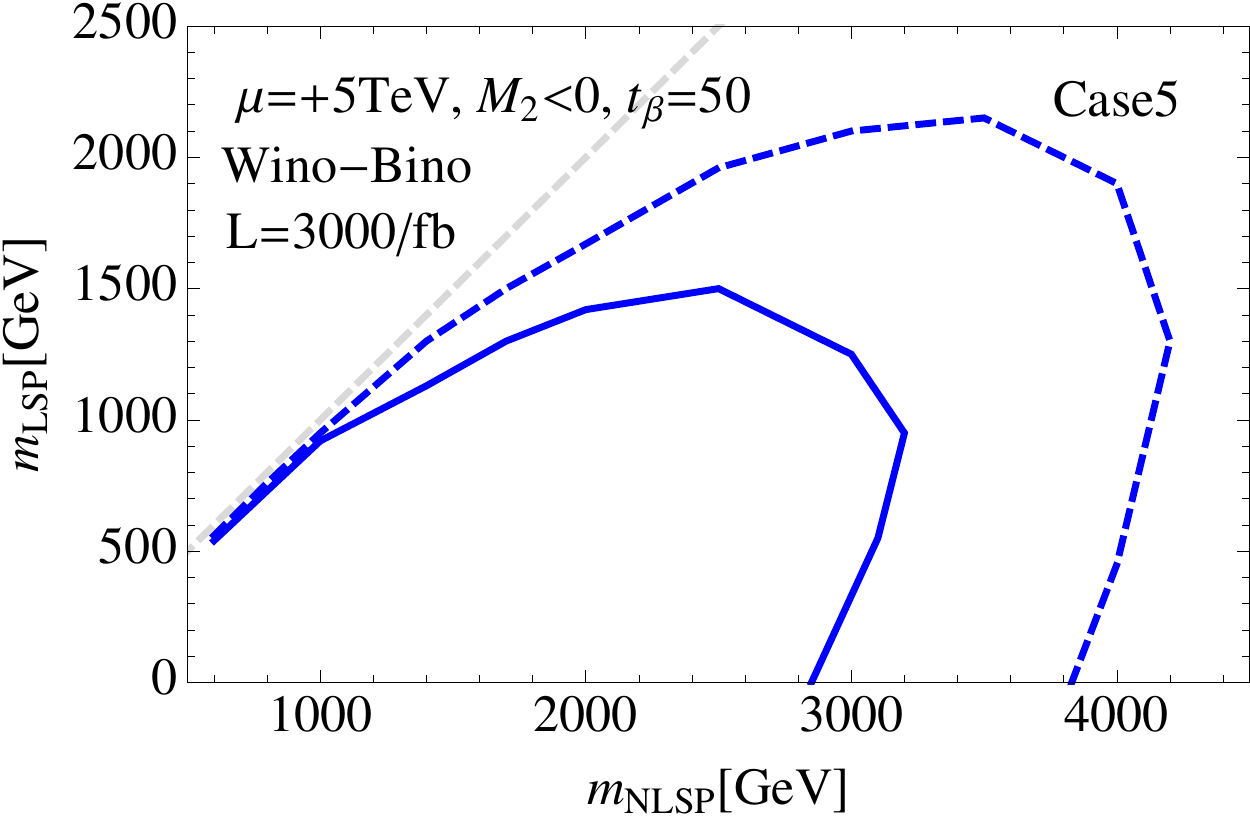}
\includegraphics[width=0.49\textwidth]{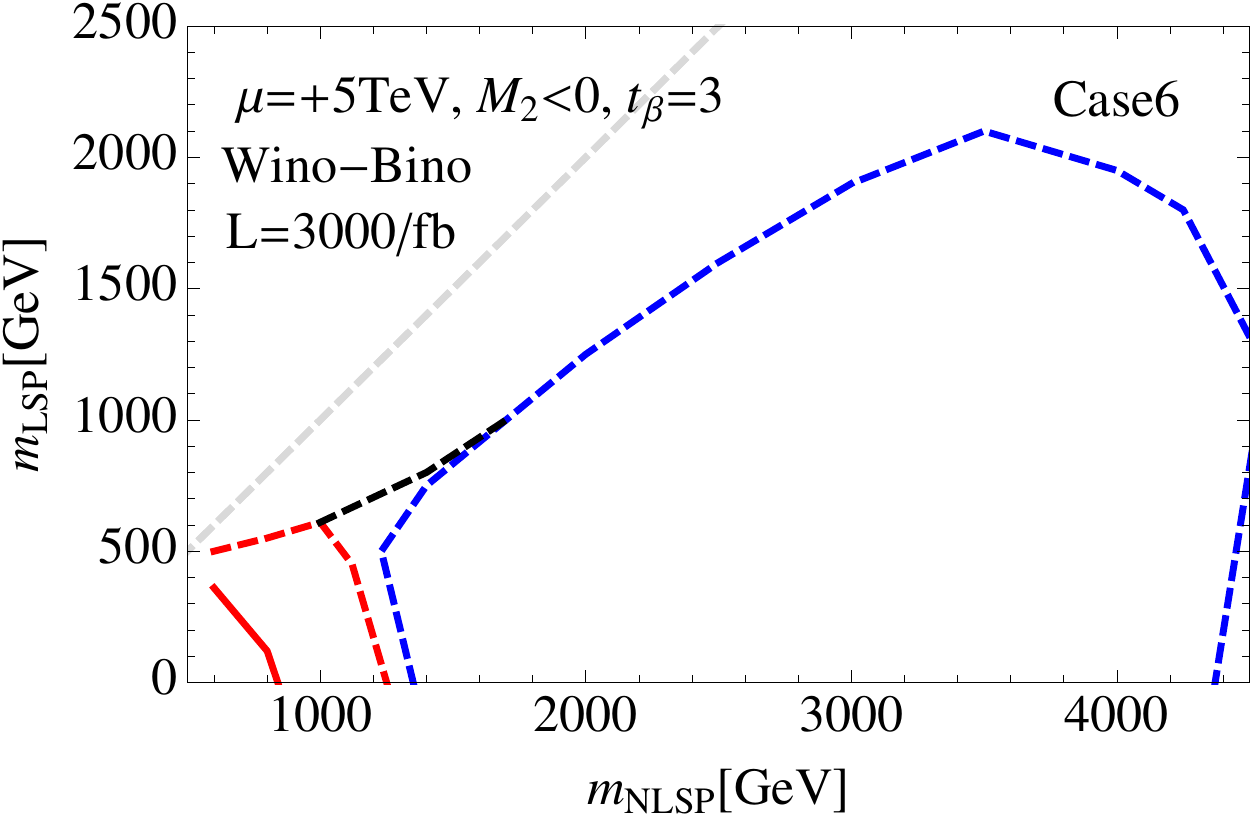}
\caption{The $3\ell$ results for the Wino-NLSP and Bino-LSP. $5\sigma$ discovery reach (solid) and 95\%\,CL exclusion limit (dashed) are shown. We show individual results from $WZ$ (blue) and $Wh$ (red) channels in separate colors, but we also show combined results (black) when both channels contribute similarly. Parameters in each case are also tabulated in Table~\ref{tab:WB-cases}.}
\label{fig:WB-3l-reach}
\end{figure}

\begin{figure}[t] 
\includegraphics[width=0.49\textwidth]{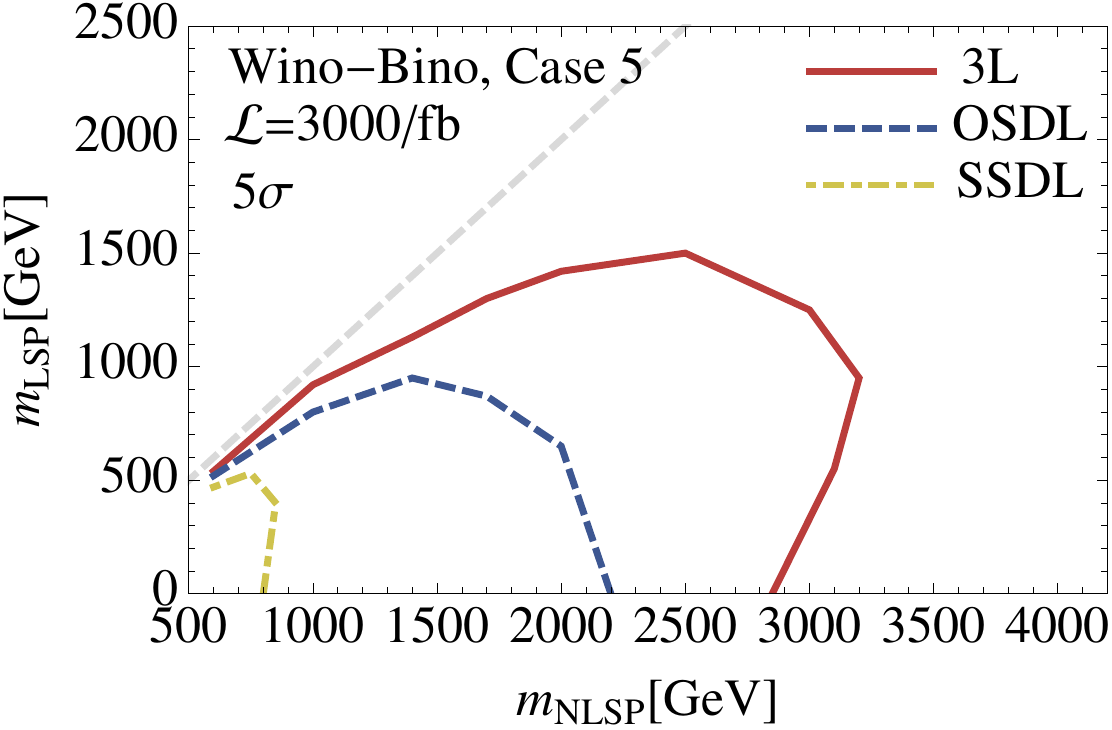}
\includegraphics[width=0.49\textwidth]{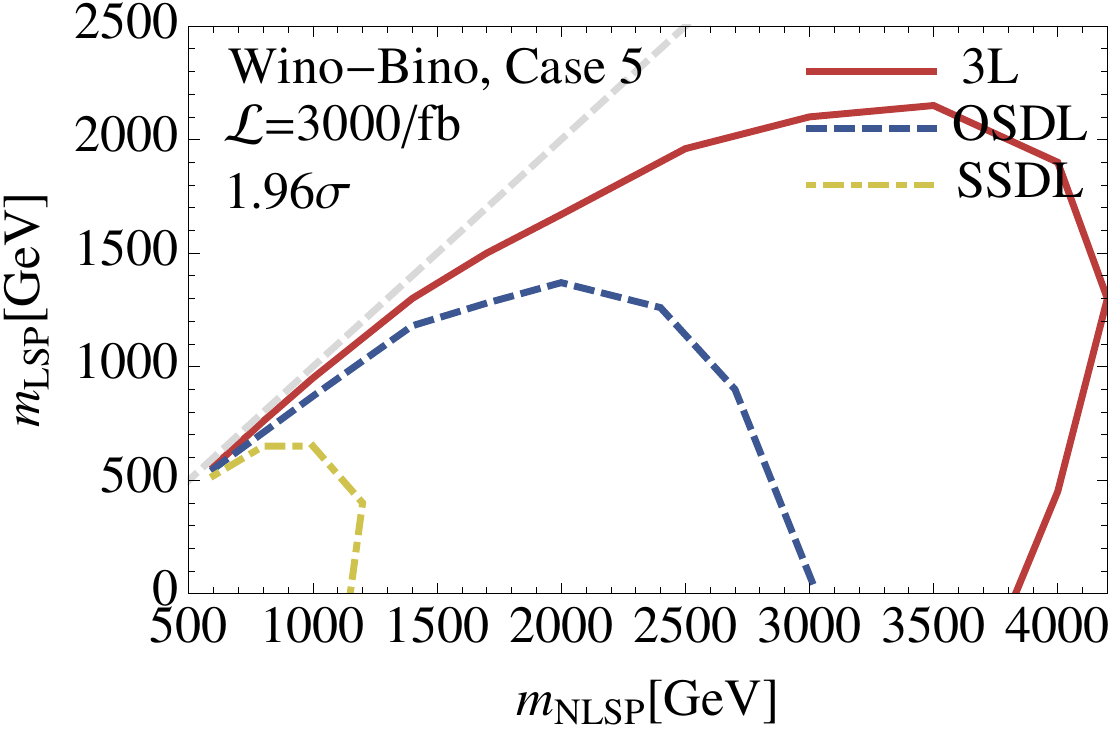}
\caption{Same as \Fig{fig:WH-reach} for the Wino-NLSP and Bino-LSP, Case 5 benchmark scenario.}
\label{fig:WB-2l-reach}
\end{figure}

\medskip

We now discuss various features of the $3\ell$ results in \Fig{fig:WB-3l-reach}. 
We first collect them here, and explain them analytically below.
\begin{enumerate}

\item Flatness of the reach curves: For Case 1 and 2, the reach curves are relatively flat, whereas wider regions of mild- or small-gap can be probed for Case 5 and 6.

\item Shape of the reach curves at high mass end: Case 4, 5 and 6 show the reach curves bending backward in the NLSP mass, since the $WZ$ reach is maximum at some non-zero LSP mass. On the other hand, Cases 1, 2 and 3 show more typical curve shapes reaching the highest mass with massless LSPs. 

\item Channel dominance: In Case 3, only the single channel $Wh$ (red curve in the figure) is contributing to the reach in the whole parameter space shown. On the other hand, in Case 2 and 5, only $WZ$ (blue curve in the figure) is the dominant channel. 
Differently, in Case 1, 4 and 6, there is a transition from $Wh$ to $WZ$ dominance: 
  the $Wh$ channel is best at small NLSP masses but the $WZ$ takes over in the high mass region. 

\end{enumerate}

In the parameter space with well-separated electroweakino masses, the relative branching ratios into the $Z$ and Higgs bosons can be approximated using the Goldstone equivalence theorem~\cite{Jung:2014bda}
\beq
\frac{\Gamma(\widetilde{W}^0 \to \widetilde{B}^0 Z)}{\Gamma(\widetilde{W}^0 \to \widetilde{B}^0 h)}  \simeq 
\frac{ | (s_\beta N_{14} + c_\beta N_{13}) - t_W ( s_\beta N_{24} + c_\beta N_{23}) |^2 \, ( 1-2r) }{ | (s_\beta N_{14} - c_\beta N_{13}) - t_W ( s_\beta N_{24} - c_\beta N_{23}) |^2 \, ( 1+ 2r)},
\label{eq:width-approx1} \eeq
valid in the approximation $|M_1|\ll M_2$ and 
where $r \equiv m_{\widetilde{B}}/m_{\widetilde{W}} \simeq M_1/M_2$ can either be positive or negative depending on the relative sign of parameters. 
The mixing angles $N_{ij}$ are approximated in the heavy Higgsino limit by \cite{Gunion:1987yh}
\beq
\left( \bmat N_{13} & N_{14} \\ N_{23} & N_{24} \emat \right) \, \simeq \, \left( \bmat \displaystyle \frac{ m_Z s_W c_\beta ( M_1 + \mu t_\beta )}{ (\mu^2 - M_1^2 )} & \displaystyle -\frac{m_Z s_W c_\beta ( \mu + M_1 t_\beta )}{ (\mu^2 - M_1^2 )}  \\[0.8em]  \displaystyle -\frac{m_Z c_W c_\beta( M_2 + \mu t_\beta )}{ (\mu^2 - M_2^2 )} & \displaystyle \frac{ m_Z c_W c_\beta( \mu + M_2 t_\beta )}{ (\mu^2 - M_2^2 ) } \emat \right),
\label{eq:mixing-WinoBino}\eeq
where $N_{13}$ ($N_{14}$) are the Bino-like mass eigenstate $\tilde H_d^0$ ($\tilde H_u^0$) components, and $N_{23}$ ($N_{24}$) are the Wino-like mass eigenstate $\tilde H_d^0$ ($\tilde H_u^0$) components.
By plugging \Eq{eq:mixing-WinoBino} into \Eq{eq:width-approx1} and taking the limit $M_1 \to 0$, we arrive at
\beq
\frac{\Gamma(\widetilde{W}^0 \to \widetilde{B}^0 Z)}{\Gamma(\widetilde{W}^0 \to \widetilde{B}^0 h)} \, \simeq \, \frac{ (M_2 \mu c_{2\beta})^2 }{ ( (2\mu^2-M_2^2) s_{2\beta} + M_2 \mu )^2} \, \simeq \, \frac{ (M_2 c_{2\beta})^2 }{ ( 2\mu s_{2\beta} + M_2 )^2},
\label{eq:width-approx2} \eeq
where we used $|\mu|>|M_2|$ in the second approximation. 
This relation keeps all the leading dependences on relative signs between $\mu$ and $M_2$ that can lead to important cancellations. The approximation is valid up to ${\cal O}(M_2^2 / \mu^2)$ terms. If we further assume that $2|\mu| s_{2\beta} \gg |M_2|$, the ratio gets the familiar form 
\beq
\frac{\Gamma(\widetilde{W}^0 \to \widetilde{B}^0 Z)}{\Gamma(\widetilde{W}^0 \to \widetilde{B}^0 h)} \sim \frac{M_2^2}{4 \mu^2 } \frac{1}{ t_{2\beta}^2}  \qquad  {\rm for} \quad 2 |\mu| s_{2\beta} \gg |M_2| \gg M_1.
\label{eq:width-approx3} \eeq
In this limit, it is evident that
the Wino dominantly decays to Binos via Higgs bosons rather than $Z$ bosons~\cite{Gunion:1987yh,Baer:2012ts}. The statement is further supported by the observation that the Wino-Bino-Higgs coupling needs only one small mixing insertion while the Wino-Bino-$Z$ coupling needs two. This statement is generally true if Higgsinos are very heavy. However, in a large part of the parameter space with mildly heavy Higgsinos, the condition $2|\mu| s_{2\beta} \gg |M_2|$ is not satisfied, and the Goldstone equivalence theorem inherently relates the Wino-Bino-$Z$ process with the Wino-Bino-Goldstone process which needs only one mixing insertion~\cite{Jung:2014bda}. This is especially true when the $\mu$ and $M_2$ have opposite signs and lead to a partial cancellation in the denominator of \Eq{eq:width-approx2}. They can lead to the dominance of Wino decays to $Z$ bosons.

If we restore the leading dependence on $M_1$, \Eq{eq:width-approx2} becomes (still in the limit of $|\mu| \gg |M_1|, |M_2|$)
\beq
\frac{\Gamma(\widetilde{W}^0 \to \widetilde{B}^0 Z)}{\Gamma(\widetilde{W}^0 \to \widetilde{B}^0 h)} \, \simeq \, \frac{  c_{2\beta}^2 (M_2 + M_1)^2 (1-2r) }{ ( 2\mu s_{2\beta} + M_2 +M_1 )^2 (1 +2r)} \, \simeq \, \frac{  c_{2\beta}^2 M_2^2  }{ ( 2\mu s_{2\beta} + M_2 +M_1 )^2 (1 +2M_1/M_2)}.
\label{eq:width-approx4}
\eeq
This expression keeps all the leading dependences on the relative signs of mass parameters. The approximation is valid up to ${\cal O}(M_1^2/\mu^2, M_2^2/\mu^2, M_1^2/M_2^2)$. 

All the features discussed above in the $3\ell$  reach can be understood from these analytic approximate expressions. We also show BRs of NLSP Winos in \Fig{fig:WBbr} to help understanding the results. 
\begin{figure}[t] 
\includegraphics[width=0.45\textwidth]{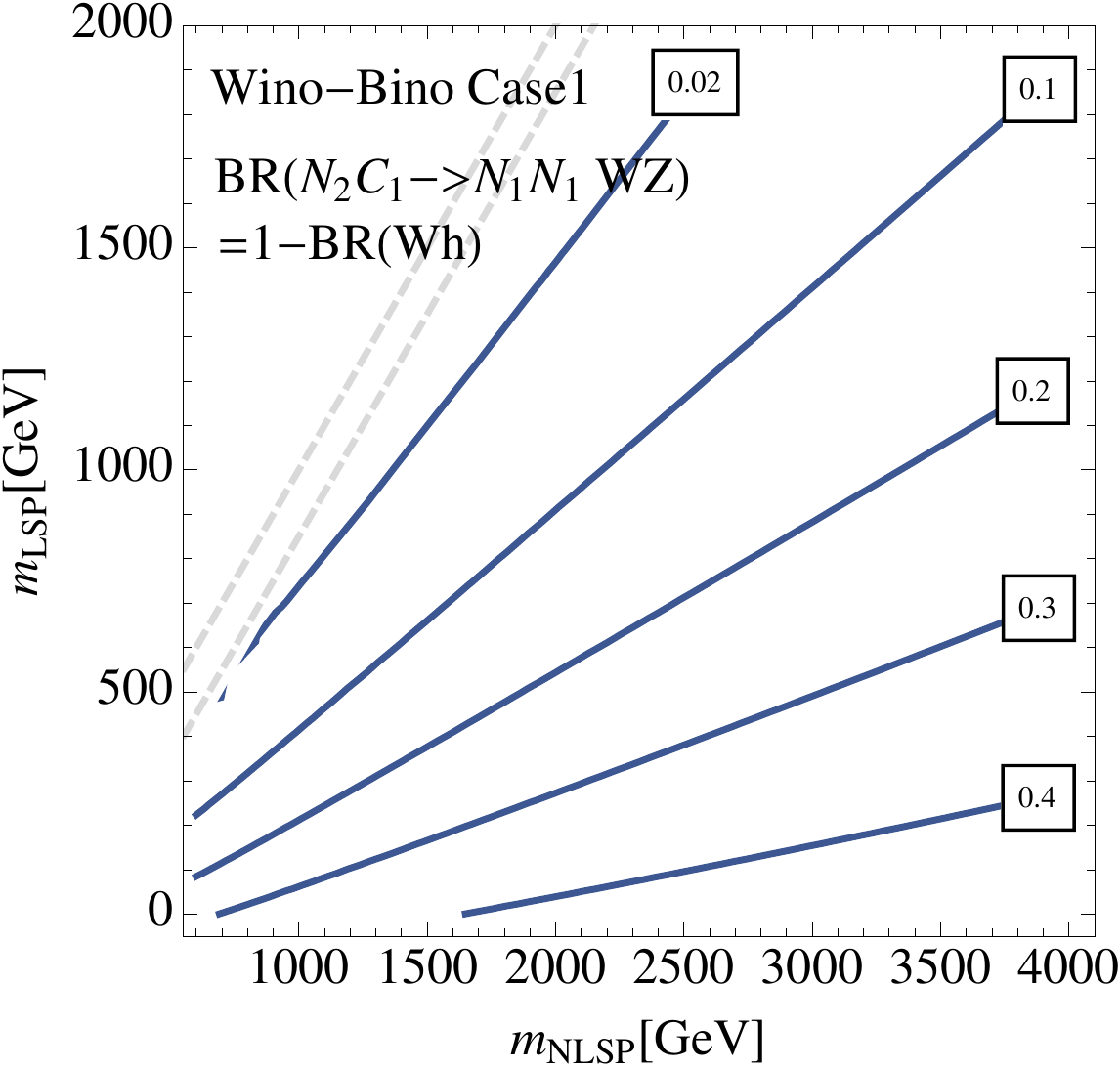}
\includegraphics[width=0.45\textwidth]{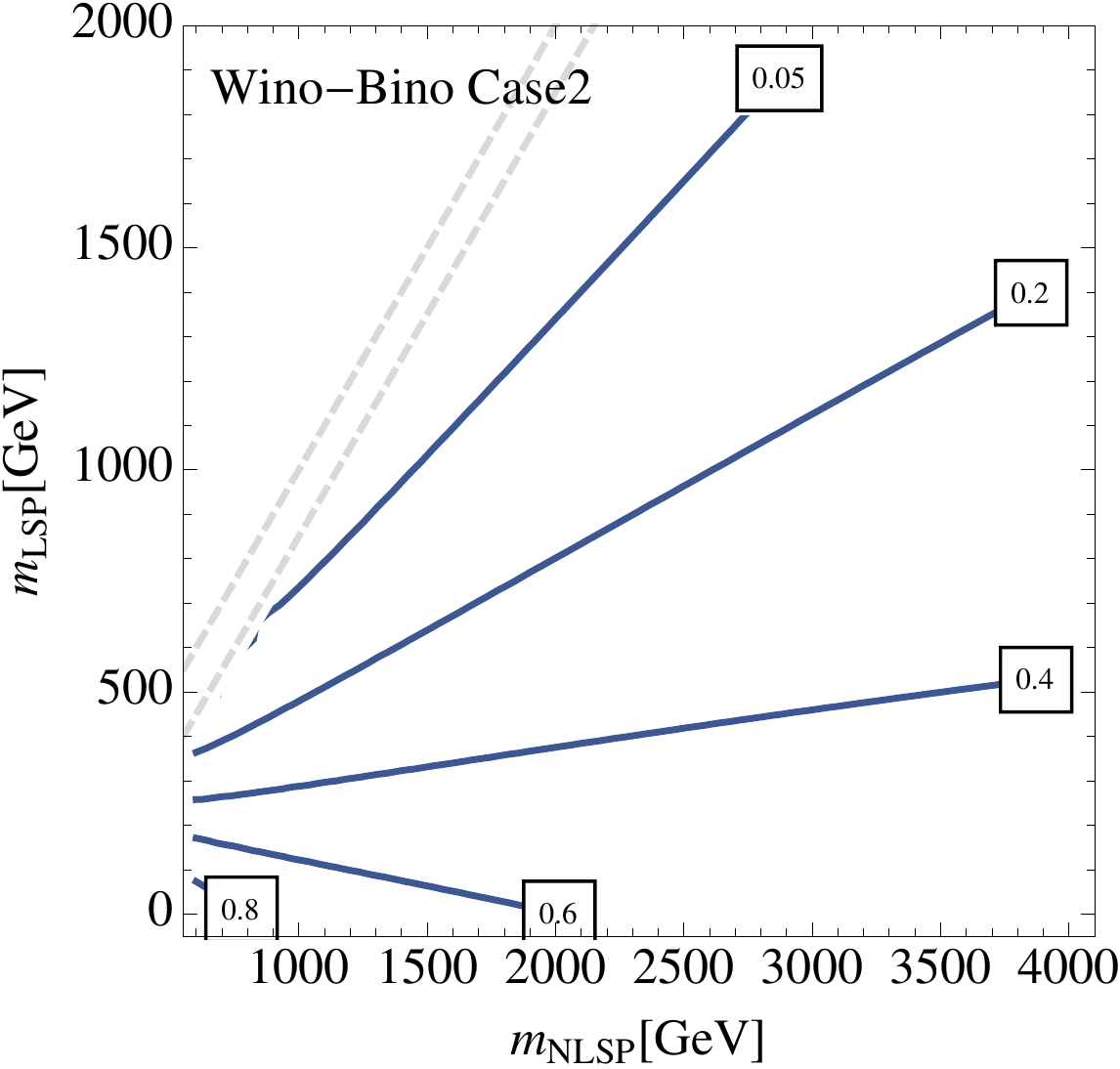}\\
\includegraphics[width=0.45\textwidth]{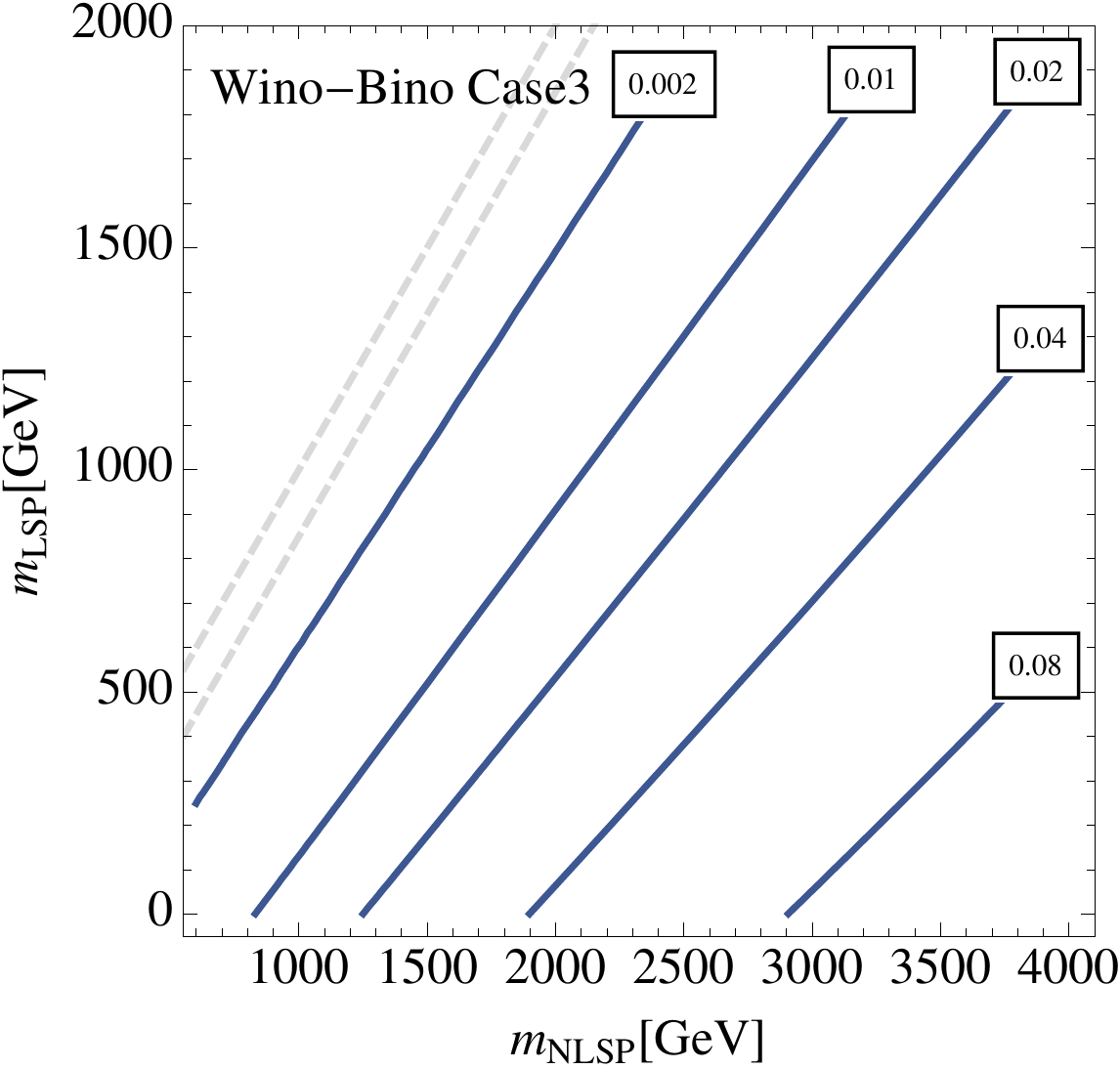}
\includegraphics[width=0.45\textwidth]{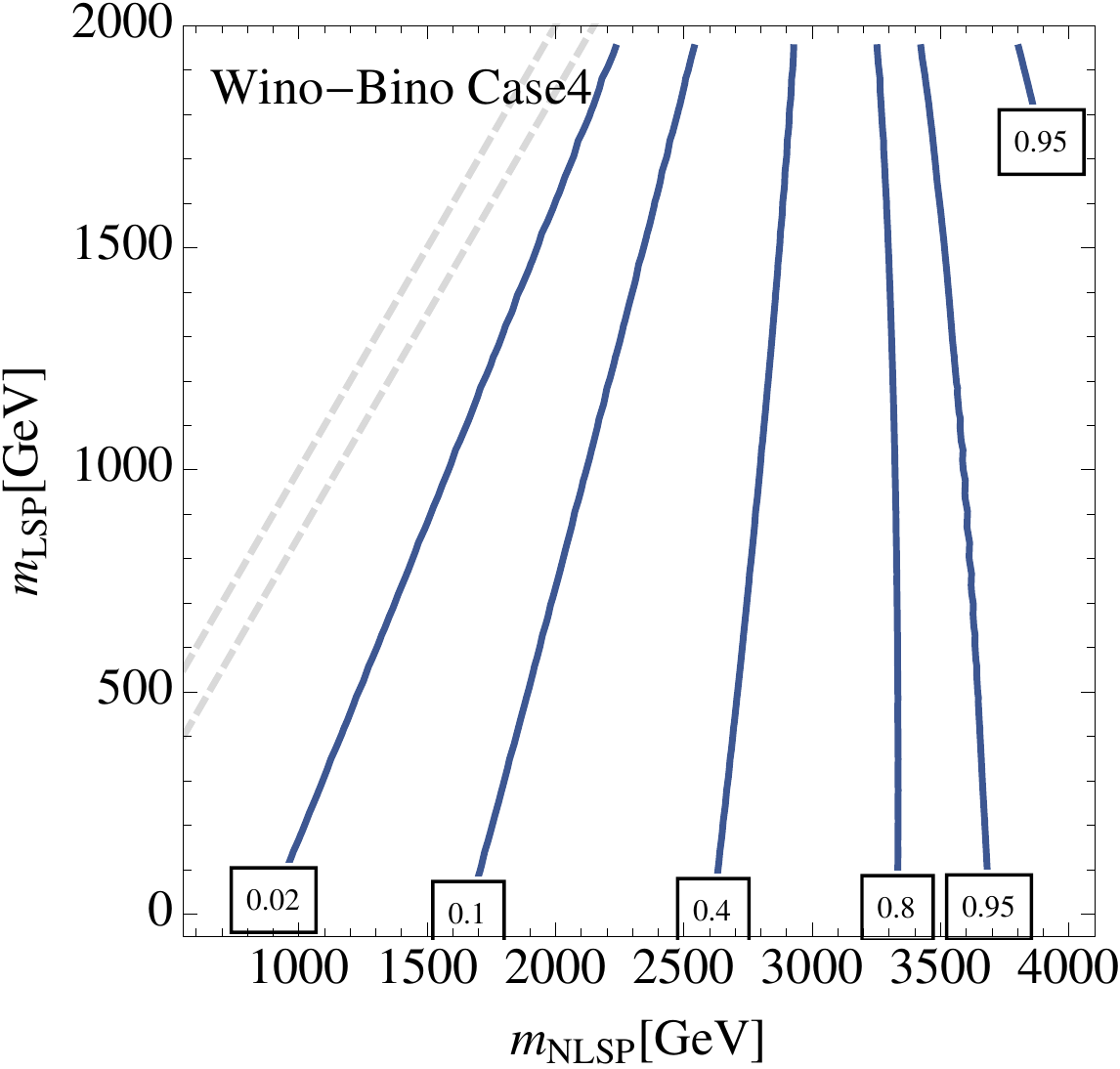}\\
\includegraphics[width=0.45\textwidth]{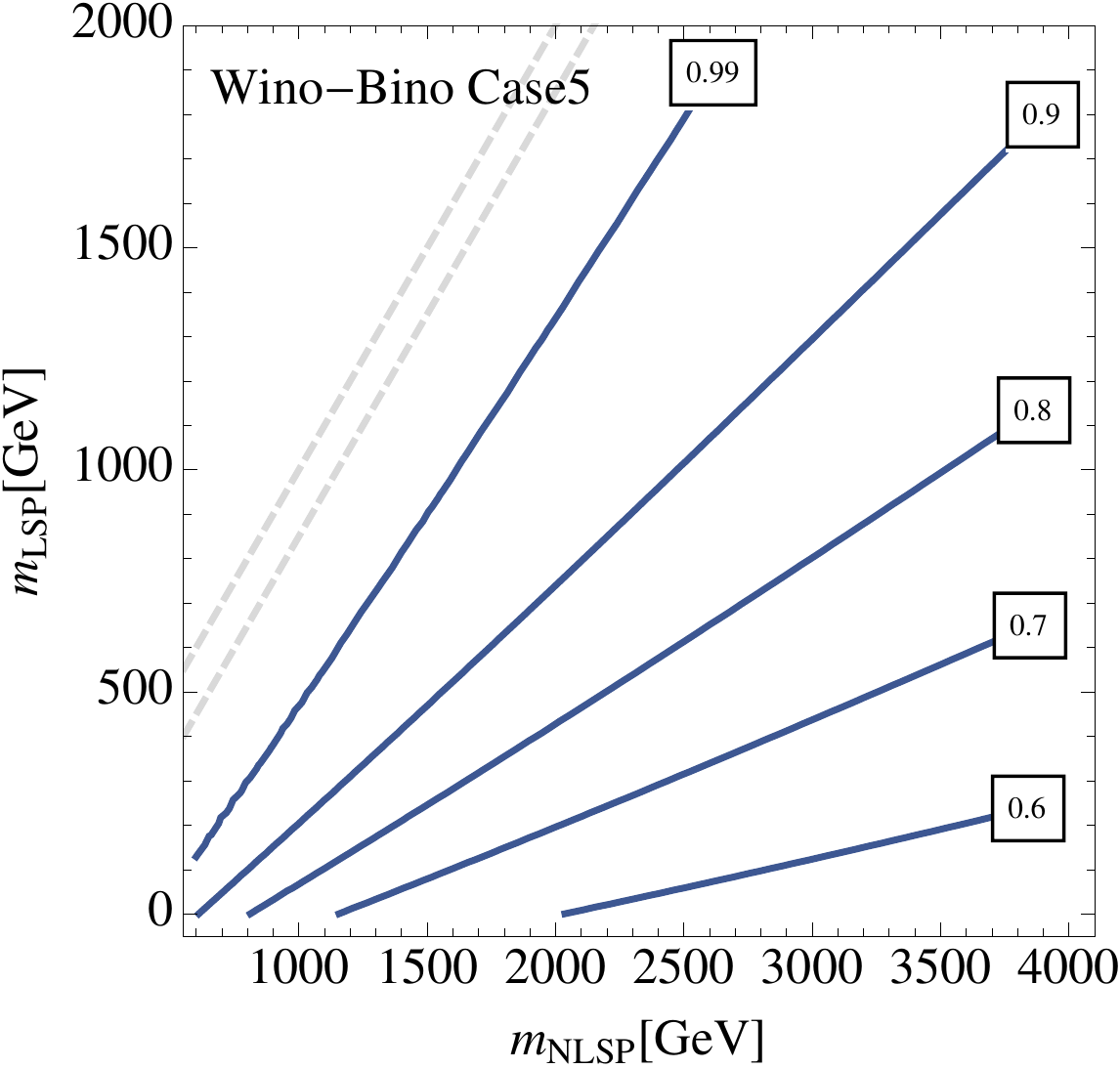}
\includegraphics[width=0.45\textwidth]{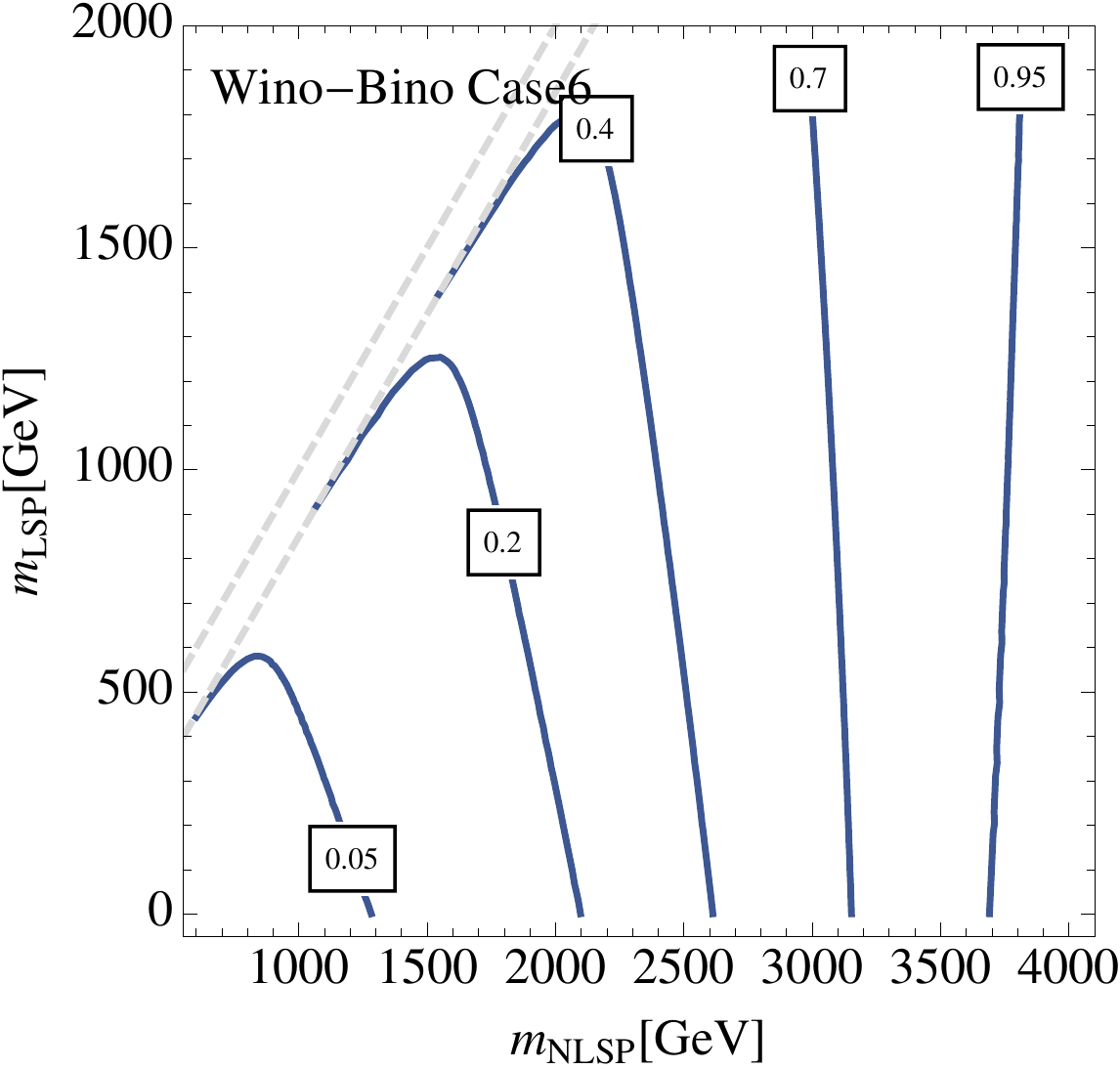}
\caption{The Wino NLSP branching ratio relevant to the $3\ell$ search, BR($N_2 C_1 \to N_1 N_1 WZ$). The two dashed lines are at $m_{\rm{NLSP}}-m_{\rm{LSP}} = 0$ and $m_{\rm{NLSP}}-m_{\rm{LSP}} = m_h$ between which we always have BR($N_2 \to N_1 Z) \approx 1$.
}
\label{fig:WBbr}
\end{figure}

\bei

\item Let us consider the $M_1 \to 0$ limit. Only the relative sign$(M_2 \mu)$ and $t_\beta$ are relevant (see \Eq{eq:width-approx2}).  Case 2 and 5 differ only by the sign($M_2 M_1$), and thus they have the same mass reach along the massless LSP line ($M_1=0$). Likewise, Case 4 and 6 have the same reach with massless LSPs.

\item The flatness of the reach curve is dictated by the sign($M_2 M_1$). From \Eq{eq:width-approx4} we see that the sign determines how the branching ratio changes with the mass-gap. As $M_1$ approaches $|M_2|$, the $Z$ mode branching ratio becomes larger if  sign($M_2 M_1)<0$; thus, the reach curve extends towards the small-gap region covering a wider parameter space. Otherwise, the reach curve tends to be flatter. Case 1, 2 vs. 5 as well as case 4 vs. 6 can be compared to observe this behavior.

\item The shape of the reach curve at the high mass end is also explained by the sign($M_2 M_1)$. As shown in Eq.~\ref{eq:width-approx4}, if sign($M_2 M_1)<0$, the branching ratio to $WZ$ becomes larger as we raise the LSP mass, resulting in better reach. 
Of course, this effect is limited if the mass gap is too small. As usual, this compressed region suffers from low efficiencies and therefore worse sensitivities.

\item 
The mode $N_2 \to N_1 h$ is dominant at small $t_\beta$ as most clearly shown in \Eq{eq:width-approx3}. It is especially dominant when sign($\mu M_2)>0$, where no cancellation in the Higgs partial width is possible, as shown in \Eq{eq:width-approx2}. If the sign($\mu M_2)<0$, even a small value of $\tan\beta$ does not guarantee the dominance of the $h$ mode. This behavior can be seen by comparing Case 3 with sign($\mu M_2)>0$ to Case 4 and 6 with sign($\mu M_2)<0$.

\item The transition  of $Wh$ channel dominance to $WZ$ channel dominance is generically dictated by the suppression factor $M_2^2/\mu^2$ in \Eq{eq:width-approx3}. As $M_2$ grows, the $WZ$ signature becomes relatively more important. The behavior generally appears in the high mass region $M_2 \gtrsim 3$\,TeV, which is not far from the value we chose for the $\mu$ parameter: $|\mu|=5$ TeV.
\eei

\medskip

What if Higgsinos are much heavier than 5 TeV, as assumed in our Figs.~\ref{fig:WB-3l-reach},~\ref{fig:WB-2l-reach}? If Higgsinos are heavy enough to satisfy  \Eq{eq:width-approx3} reasonably well,
 Higgs channels always dominate and the $3\ell$ reach becomes weaker. The reach will be rather low, similar to that of Case 3. On the other hand, the OSDL searches are not affected by the exact choice of the $\mu$ parameter, as long as $|\mu|\gg|M_2|$, so that the chargino is mainly Wino-like, 
 because the relevant BR, BR($C_1 C_1 \to N_1 N_1 WW$), is always close to 100\%. For this reason, the OSDL channel can become the leading discovery channel and a hint for a spectrum with very heavy Higgsinos.

%%%
\subsection{Comparison with Nearby Gluino Reach} \label{sec:comparison}

The gluino pair is usually a better discovery channel if gluinos are not too much heavier than electroweakinos. 
It is interesting to identify in which circumstances heavy gluinos are more difficult to search for than electroweakino NLSP pairs studied here.

Gluino pairs can be excluded at a 100 TeV collider with 3/ab when gluinos are lighter than about 14 TeV~\cite{Cohen:2013xda,Jung:2013zya}. As long as gluinos are lighter than about 12--13 TeV, up to 4 TeV LSPs can be excluded regardless of gluino masses. Meanwhile, as we have shown in our paper, only up to 1--2 TeV LSPs can be excluded from multi-lepton NLSP searches. Thus, if the gluino is lighter than 12--13 TeV, it is generally an earlier discovery channel.

In the majority of SUSY models~\cite{Choi:2007ka}, gaugino masses are predicted to have order-one ratios of each other, which means that gluinos are typically not much heavier than the other gauginos. In such scenarios, if the gluino is out of the reach of a 100 TeV collider, $> 13 $ TeV, can we still have prospects of discovering the lighter electroweakinos? As examples, we consider a couple of well known SUSY breaking models.

With the mSUGRA relation,  $M_1 : M_2 : M_3 \simeq 1: 2: 6$, the 13 TeV gluino implies a 2 TeV Bino and a 4.2 TeV Wino. If Higgsinos are LSPs, lighter than the 2 TeV Binos, no exclusion is expected from Bino NLSP production nor Wino NNLSP productions (see \Fig{fig:WH-reach}). No exclusion is also expected  when the Higgsino is the NLSP with mass between 2 and 4.2 TeV (see \Fig{fig:HB-reach}).

The AMSB scenario is more interesting, as it predicts a larger gluino-wino mass splitting.
The relation, $M_1: M_2 : M_3 \simeq 3: 1: 8$ -- renormalized at 2 TeV by including two-loop gauge coupling runnings and one-loop threshold corrections~\cite{Gupta:2012gu,Jung:2013zya} -- implies that Winos can be as light as $1.6$ TeV (while the 5 TeV Bino is irrelevantly heavy) when the gluino is above 13 GeV. If Higgsinos are lighter than Winos, the 1.6 TeV Wino NLSPs can probe up to 1.2 TeV Higgsino LSPs (see \Fig{fig:WH-reach}). If Higgsinos are NLSPs, however, a 1.6 TeV Wino LSP is not expected to be excluded from Higgsino NLSP pair productions (see \Fig{fig:HW-reach}).

In all, there are chances that multi-lepton searches of  NLSPs can lead to an earlier discovery of SUSY than direct gluino searches, for example, in the AMSB scenario.

%%%%%%%
\section{Detector Optimization Issues} \label{sec:detector}

In this section, we briefly discuss possible detector developments that can improve and optimize our multilepton searches.

The pair of leptons coming from heavy electroweakino decays, ${\rm{NLSP}}\to {\rm{LSP}} Z, \, Z\to\ell\ell$, will be collimated at a 100 TeV collider, if the mass splitting between the NLSP and the LSP is sizable. In \Fig{fig:mindrll}, we show distributions of minimum angular separation between any two leptons from the $3\ell$ and OSDL 
signal events. Typical angular separation between the pair  is $\Delta R \sim m_{\rm Z }/ 2 m_{NLSP } $, which can be smaller than the lepton separation criteria we use in our analysis, $\Delta R>0.05$. In that circumstance the two leptons will be reconstructed as a single jet. This can degrade the performance of multi-lepton searches. 

\begin{figure}[t] 
\includegraphics[width=0.99\textwidth]{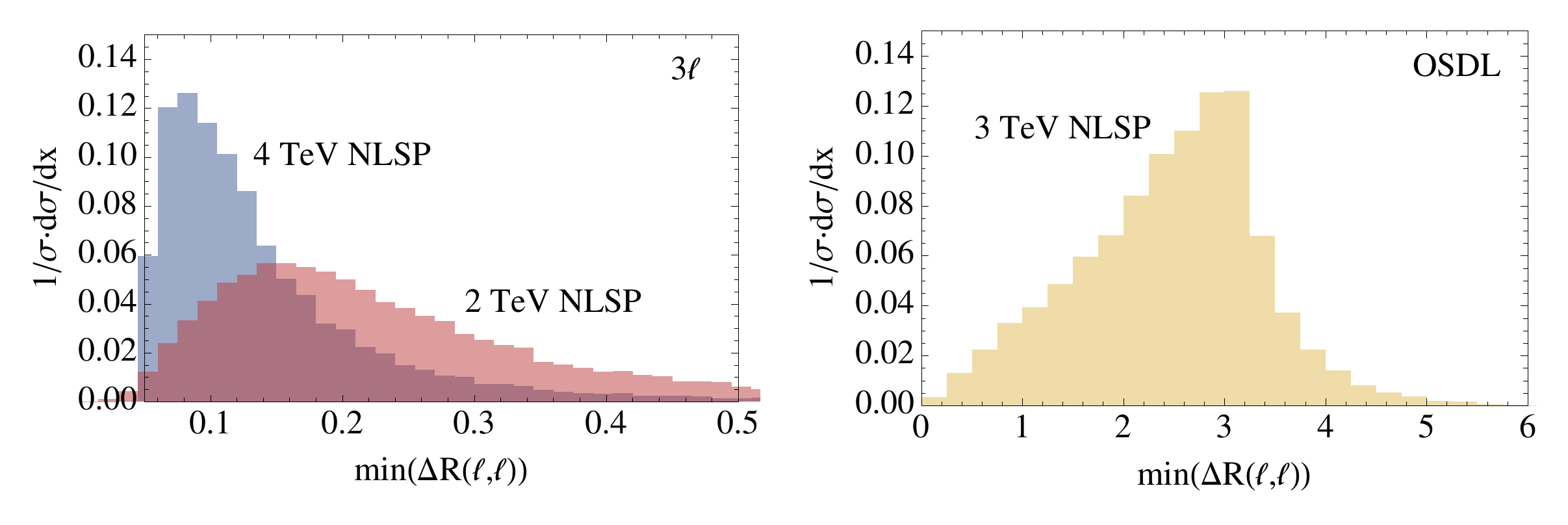}
\caption{Minimum lepton separation angle, $\Delta R(\ell,\ell)$, in the $3\ell$ {(left panel)} and OSDL {(right panel)} searches for massless LSP.}
\label{fig:mindrll}
\end{figure}

\begin{figure}[t] 
\includegraphics[width=0.49\textwidth]{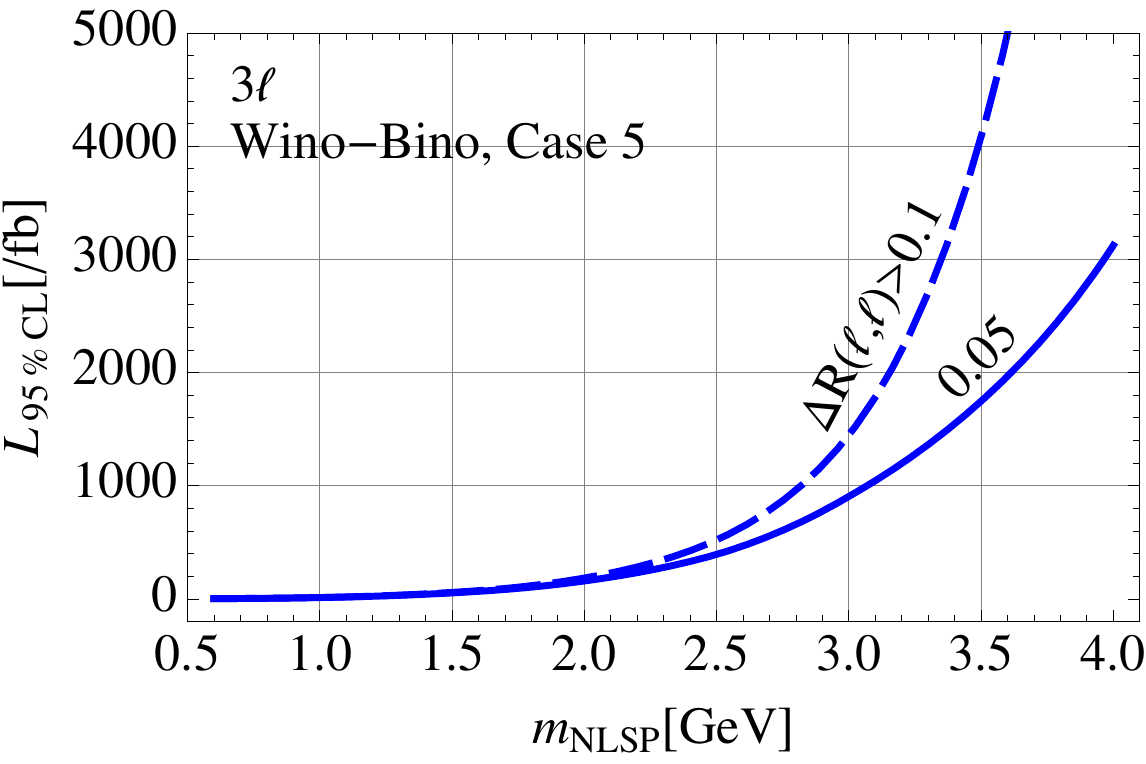}
\includegraphics[width=0.49\textwidth]{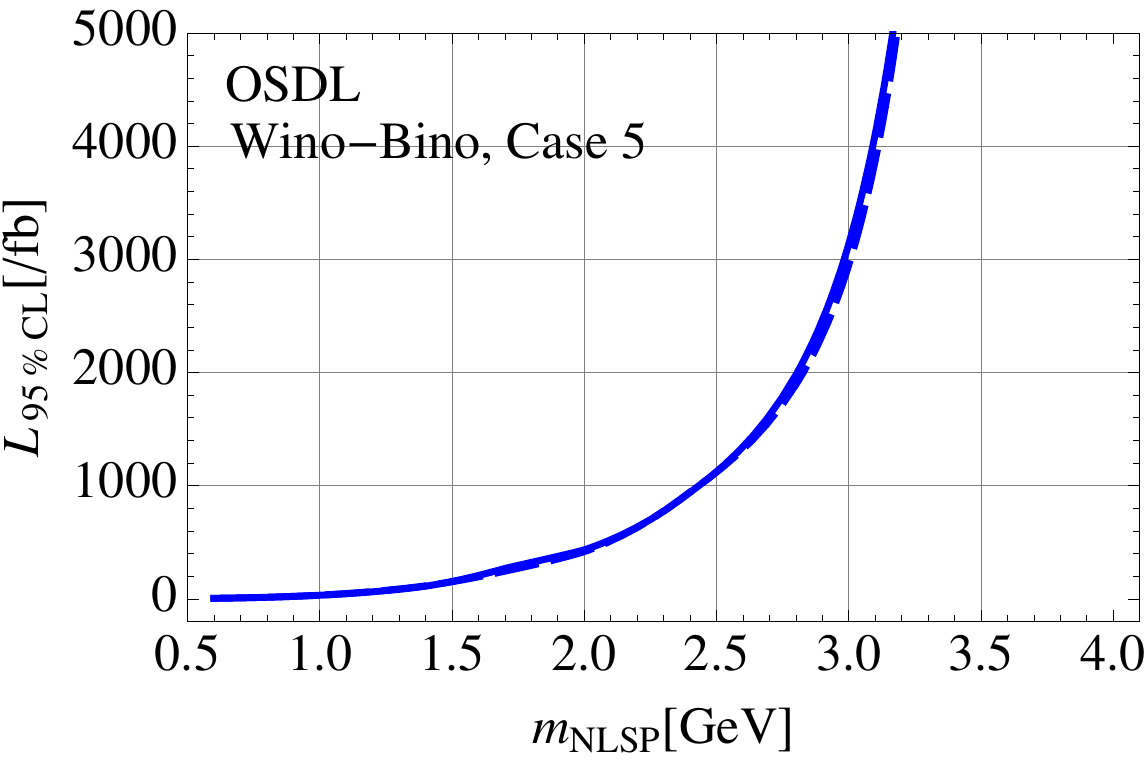}
\caption{By varying the lepton separation criterion $\Delta R>$ 0.1 (dashed) and 0.05 (solid), we show the luminosity needed for 95\%CL limit in the $3\ell$ ({left panel}) and the OSDL ({right panel}) searches. The Wino-Bino Case 5 with massless LSP is used -- solid lines correspond to \Fig{fig:WB-2l-reach} results. 
}
\label{fig:lum-collimation}
\end{figure}

We illustrate this issue in the left panel of \Fig{fig:lum-collimation}, where we show the dependence of the $3\ell$ results on the lepton separation criterion.
In particular, we present the luminosity needed for the 95\% CL exclusion with separation criterion varied between $\Delta R>0.1$ and 0.05.
With the $\Delta R>0.1$ criterion, the degradation of the $3\ell$ reach compared to reach obtained with $\Delta R>0.05$ begins to appear 
at NLSP masses at around 
2.5--3.0 TeV with about 1/ab of data. For example, the luminosity needed to probe a 3.5 TeV Wino would be almost doubled with the separation requirement $\Delta R>0.1$, compared to the one with $\Delta R>0.05$.

We also verify that leptons are usually well separated in the OSDL (and SSDL) channels, since they are mainly from different $W$ bosons in the $W^+W^-$ ($W^\pm W^\pm$) channel. Therefore, the reach is not significantly affected by the ability of lepton separation technique, as demonstrated in the right panel of \Fig{fig:lum-collimation}.

\begin{figure}[t] 
\includegraphics[width=1.0\textwidth]{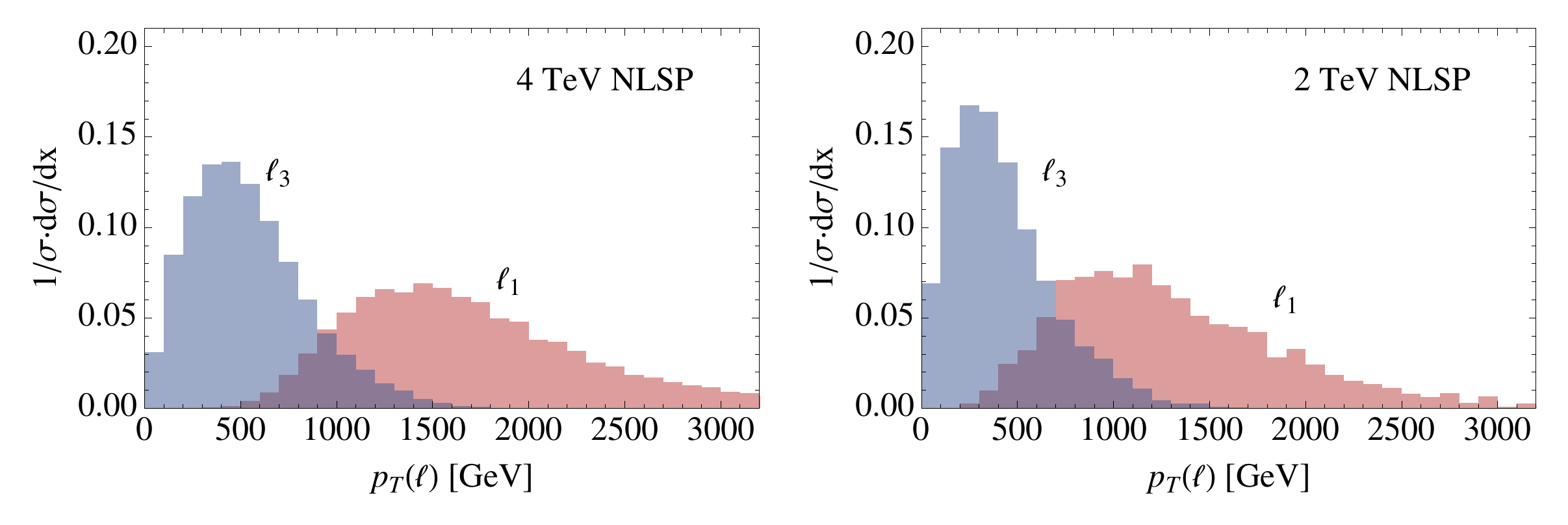}
\caption{Lepton $p_T$ distributions from 4 TeV {(left panel)} and 2 TeV {(right panel)} NLSP pair decays 
giving the $3\ell$ signature. Leading lepton (red) and 3rd lepton (blue) are shown. All discovery cuts are applied.}
\label{fig:leppt}
\end{figure}

As shown in \Fig{fig:leppt}, leading leptons typically have TeV-scale energies. The identifications of such highly boosted lepton's flavor and charge are additional potential challenges that should be addressed at future colliders. The SSDL search channel can be particularly affected by this issue. Abundant electromagnetic radiations off of energetic muons may make them less efficiently tagged than electrons. And detector magnetic fields may not be strong enough to bend fast-moving charged leptons enough.

Finally, a 100 TeV collider will be an environment full of hadronic activity. Lepton-jet isolation techniques can thus be important. As an example, if we relax the isolation criteria to allow soft jets nearby a  lepton (specifically, if a nearby jet is softer than the lepton, they are separately and properly reconstructed), we can have up to  30\% more signal samples. Such intrinsic uncertainty may reside in our analysis of the future high-energy collider, and more careful assessment will be useful when detector performances become known.

%%%%%%%%
\section{Conclusions}\label{sec:concl}

In this paper, we have studied the discovery prospects of multi-lepton searches of electroweakinos at a 100 TeV proton-proton collider. In particular, we have studied the $3\ell$, 
opposite sign di-lepton (OSDL) and same sign di-lepton (SSDL) final states and considered various possible NLSP-LSP combinations in the MSSM. 
We summarize our results in Table~\ref{tab:discovery-exclusion}.
\begin{table}[h!] \centering
\begin{tabular}{ccc}
\hline \hline
                     & $5\sigma$  & $95\%$ CL \\
(NLSP, LSP) & discovery & exclusion \\
\hline
$(\tilde W,\tilde H)$ & (2.2, 0.8)\,TeV & (3.3, 1.3)\,TeV \\
$(\tilde H,\tilde W)$ & (1.5, 0.6)\,TeV & (2.6, 1.0)\,TeV \\
$(\tilde H,\tilde B)$ & (1.8, 0.7)\,TeV & (2.9, 1.1)\,TeV\\
$(\tilde W,\tilde B)$ & (3.2, 1.4)\,TeV & (4.2, 2.2)\,TeV\\
\hline\hline
\end{tabular}
\caption{Highest reaches among all multi-lepton searches for $5\sigma$ discovery and $95\%$ CL exclusion at 100\,TeV $pp$ hadron collider with $3\,{\rm ab}^{-1}$ of integrated luminosity. The numbers quoted for the Wino-Bino case are those obtained in Case 5.}
\label{tab:discovery-exclusion}\end{table}

These results represent a great improvement from the expected discovery reach at the 14 TeV LHC~\cite{ATLAS:2013hta,CMS:2013xfa}. Most notably, the whole parameter space of a Higgsino-like WIMP dark matter can be probed via Wino NLSPs if the Wino is lighter than about 3.2 TeV and not too close in mass to the Higgsino. Wino-like dark matter, on the other hand, is not fully probed in these searches as Wino DM is required to be quite heavy ($\sim 3.1$ TeV) and Higgsino NLSP production cross section is smaller.

We find that the $3\ell$ search, usually, has the highest signal reach. In this search, important parameter dependences may arise from $\tan \beta$ and the signs of gaugino and higgsino masses.
In the case of Higgsino LSPs or NLSPs, the results do not depend sensitively  on them, as implied by the Goldstone equivalence theorem and the Higgs alignment limit~\cite{Jung:2014bda}. As a result, the models with light Higginos (LSPs or NLSPs) can naturally serve as true simplified models with fixed BRs of NLSP neutralinos: BR($Z$) = BR($h$). On the other hand, if Higgsinos are heavier than Wino NLSPs and Bino LSPs, the parameter dependences introduce various features in the reach plot, as shown in \Fig{fig:WB-3l-reach} and discussed thereafter. The $3\ell$ reach is highest when the BR into the $WZ$ channel is maximal.

The OSDL search has advantages in the sense that parameter dependences are weaker and the lepton collimation issue is almost absent. When the $3\ell$ reach is limited by these factors, e.g. in the scenario with very heavy Higgsinos in which the dominant $Wh$ channel only leads to a weak reach, the OSDL channel can still provide a complementary sensitivity. 

Furthermore, the SSDL signal is relatively good in the low-mass small-gap region, where the soft lepton identification becomes difficult. We comment on the small-gap region, for which we did not perform a careful study. Hard initial state radiations plus soft leptons plus correlated large MET would efficiently probe the small-gap region with $m_{\rm{NLSP}}-m_{\rm{LSP}} \lesssim 50$ GeV~\cite{Gori:2013ala,Schwaller:2013baa}. This could also be studied with our kinematic variables, but we leave more dedicated assessments for future studies.

We have also studied when the direct electroweakino searches can offer an earlier discovery than the direct searches of gluino pairs.
In the AMSB models, light Wino NLSPs decaying to lightest Higgsino LSPs can be discovered earlier than the gluino pairs. In other models, however, the gluino pair search is generally better.

Searching for new physics at multi-TeV scales also presents new challenges. Our study highlights a few of them. First of all, the decay products, in particular the $Z$ boson, can be very boosted. Therefore, the two leptons from $Z$ decays will be collimated and may fail the conventional lepton isolation cuts. Secondly, measuring the properties of a energetic lepton with $p_{\rm T}^{\ell} > $ TeV, such as its flavor and charge, can be challenging. As we emphasize, both of these effect can significantly impact the reach. It will be important to optimize such performances in detector design and search strategies.

\bigskip
{\bf{Note Added:}} 
As this work neared completion, Ref.~\cite{Acharya:2014pua} appeared, whose scope
partially overlaps with ours. One notable difference of results is that our
$3\ell$ reach is stronger due to our smaller lepton separation
criteria. Furthermore, we have studied several scenarios in addition to just wino-higgsino, and introduced additional helpful kinematical variables and discussed their optimizations for various multilepton searches.

\vspace{8mm}
\noindent
{\it Acknowledgements.}
We thank D. Amidei and A. Barr for useful comments and B. Acharya, N. Arkani-Hamed and K. Sakurai for related conversations. The authors are grateful to the Mainz Institute for Theoretical Physics (MITP), Aspen Center for Theoretical Physics, which operates under the NSF Grant 1066293, and Center for Future High Energy Physics (CFHEP) in Beijing for their hospitalities and partial supports during the completion of this work. S.G. would like to thank the SLAC theory group for hospitality and partial support. Research at Perimeter Institute is supported by the Government of Canada through Industry Canada and by the Province of Ontario through the Ministry of Economic Development $\&$ Innovation. S.J. thanks KIAS Center for Advanced Computation for providing computing resources. S.J. is supported in part by National Research Foundation of Korea under grant 2013R1A1A2058449. L.T.W is supported by DOE grant DE- SC0003930. J.D.W is supported in part by DOE under grant DE-SC0011719.

%%%%%%%%%%%%%%%%%%%%%%
\appendix

%%%
\section{Validation Using a Simplified Model} \label{sec:valid}

\subsection{Simplified Model Results}

We validate our results using the simplified model of the Wino-NLSP and Bino-LSP. The simplified model is used so that we can compare directly with existing LHC results that use the same model. In particular, we assume  100\% branching ratios into the relevant diboson final states, so as to minimize model dependencies. For example, for the $3\ell$ analysis, it is assumed that chargino-neutralino pairs always decay to the $WZ$ channel which subsequently leads to the multi-lepton signal with its SM leptonic branching ratio.

In this Appendix, we use the results based on $M_{eff}$ and ``traditional'' variables (not $M_{eff}^\prime$ as discussed in \Sec{sec:opt}) that are also used in LHC8 analyses. 
First of all, we can approximately reproduce the existing ATLAS 8 limits using our event samples and optimization procedure. For the exclusion of 350 GeV-NLSP and massless LSP, the latest ATLAS 8 analysis needed 20.3/fb from the $3\ell$ search~\cite{Aad:2014nua}. Our estimation needs 19/fb, after adding the ATLAS systematic errors and normalizing our backgrounds to the ATLAS results. 

The ATLAS 8 result can be naively scaled up to the 100 TeV collider environment. We use the \texttt{Collider Reach} program~\cite{colliderreach} to obtain the corresponding limits at 100 TeV with certain luminosities. This naive scaling is expected to lead to a good estimation for the reach of those searches utilizing high-energy cuts much higher than the masses of particles because kinematic distributions at relevant high-energy regime are effectively independent of particle masses.

The scaled-up result is shown as the red-solid line in \Fig{fig:err}. In the following subsection, we compare this curve with the results we obtain varying several uncertainties; and we will see that they agree within reasonable uncertainties.

\subsection{Uncertainties From Unaccounted Effects}\label{sec:uncert}

\begin{figure}[t] 
\includegraphics[width=0.49\textwidth]{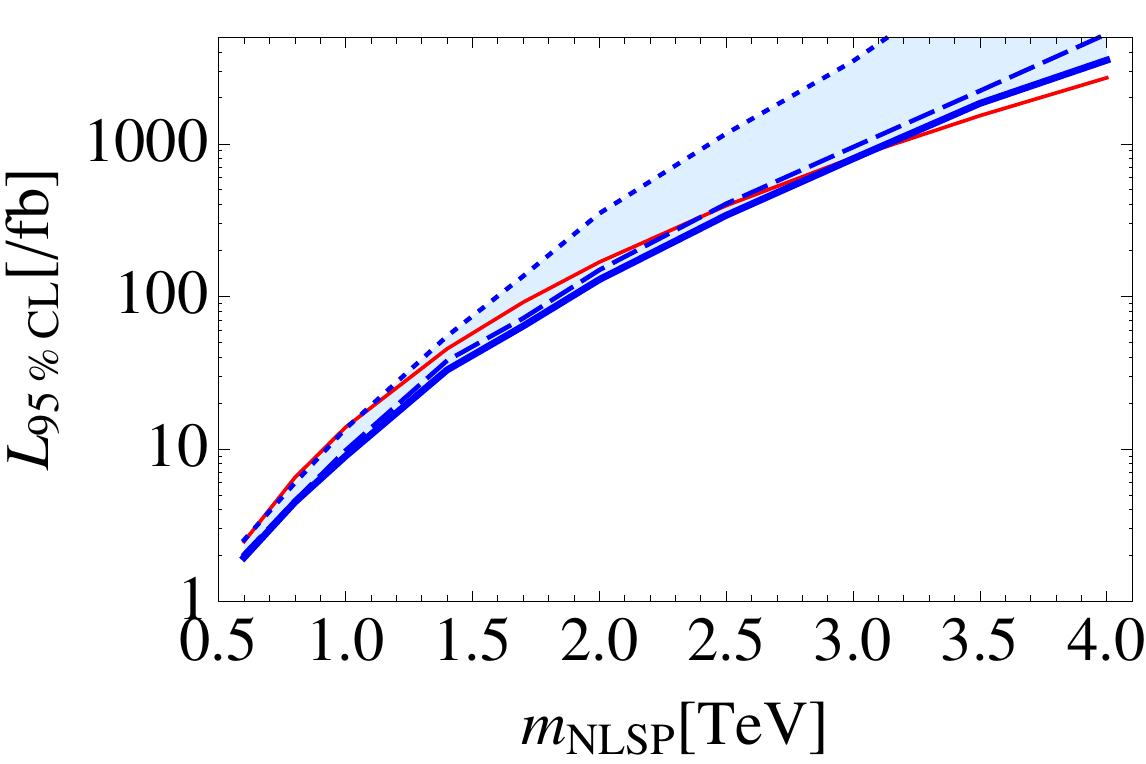}
\includegraphics[width=0.49\textwidth]{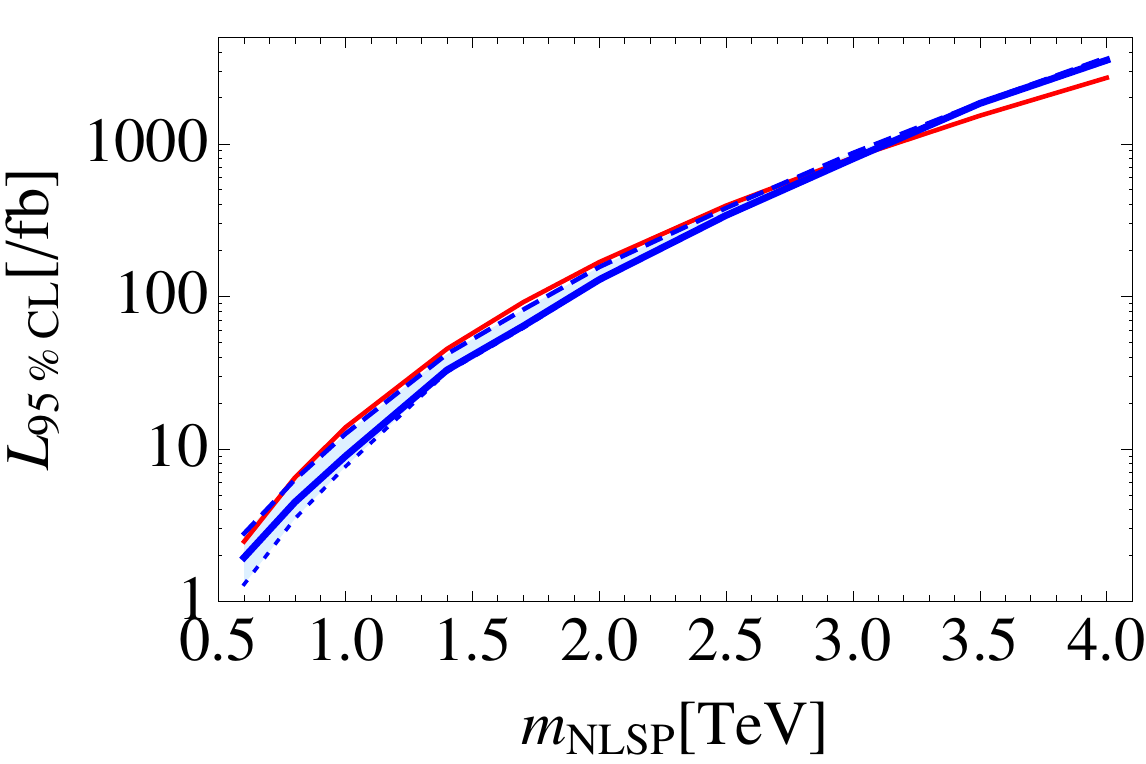}
\caption{Luminosity needed for 95\%CL limit in the $3\ell$ channel. {Left panel}: systematic error $\delta$ and background normalization $N$ are varied -- they are defined in \Eq{eq:errordef}. The blue-solid line is with no errors $\delta=0, N=1.0$ as assumed throught in this paper. The other blue lines assume $\delta=0.3, N=1.2$ (dotted), $\delta=0.05, N=1.5$ (dashed). The naively scaled LHC8 limit is also shown for comparison (red-solid). {Right panel}: minimum number of signal events after all cuts, $S$, are varied between $S>5$(blue-solid), $8$ (dashed) and 2 (dotted). In both panels, the luminosity needed for 95\%CL limit from the 3$\ell$ search is plotted using the Wino-Bino simplified model (see text for more details).}
\label{fig:err}
\end{figure}

In this section, we assess the impacts of potential systematic uncertainties, background normalization and the required  minimum number of signal events after all cuts. We parameterize the first two sources of uncertainty in the signal significance $\sigma$ as
\beq
\sigma \, \equiv \, \frac{ S}{ \sqrt{ N B + ( \delta \cdot NB )^2 } },
\label{eq:errordef}\eeq
where $S$ and $B$ are the number of signal and background events after all cuts. The systematic uncertainties are multiplicatively parameterized with $\delta$ ($\delta=0$ means no systematic errors) and the background normalization is denoted by $N$. The background normalization ($N>1$) may effectively account for subleading processes and reducible backgrounds that we did not simulate.

In the left panel of \Fig{fig:err}, we vary $N$ and $\delta$ within $N$=1--1.5 and $\delta$=0--0.3. The scaled-up ATLAS 8 result mostly falls within this uncertainty band. 
In the right panel of \Fig{fig:err}, we also vary the condition of minimum $S$ for the number of events that will be needed for the discovery. For $S$ in the range 2--8, the search capacity is not significantly modified and ATLAS 8 results mostly fall within the band. Recall that we have chosen $S>5$ throughout in this paper. We conclude that naively scaling up the LHC 8 ATLAS bound agrees reasonably well with our Simplified model results.

\subsection{Discovery Cuts Used in Tables} \label{sec:cutstable}

For the $3\ell$ results in Table.~\ref{tab:WH-cont}, \ref{tab:HW-cont}, \ref{tab:HB-cont} and \ref{tab:WB-cont} (for all models of NLSP-LSP combinations), we used, in addition to baseline cuts, 
\beq
\frac{p_T(\ell_2)}{p_T(\ell_1)} > 0.25, \quad \MET > 400\,{\rm{GeV}}, \quad M_T(W)>200\,{\rm{GeV}}, \quad M_{eff}^\prime > 900\,{\rm{GeV}}, \quad \frac{H_T(j)}{M_{eff}} < 0.25.
\eeq

For the OSDL results in Table.~\ref{tab:WH-cont}, \ref{tab:HW-cont}, \ref{tab:HB-cont} and \ref{tab:WB-cont} (for all models of NLSP-LSP combinations), we used
\beq
M_{eff}^\prime > 500\,{\rm{GeV}}, \quad M_T(W)>800\,{\rm{GeV}}, \quad \frac{\MET}{M_{eff}}>0.32, \quad \frac{p_T(\ell_2)}{p_T(\ell_1)} > 0.32.
\eeq

For the SSDL results in Table.~\ref{tab:WH-cont} and \ref{tab:HW-cont} (for wino-higgsino and higgsino-wino models), we used
\beq
M_{eff}^\prime > 500\,{\rm{GeV}}, \quad M_T(W)>720\,{\rm{GeV}}, \quad \MET>260\,{\rm{GeV}}, \quad \frac{p_T(\ell_2)}{p_T(\ell_1)} > 0.2.
\eeq

For the SSDL results in Table.~\ref{tab:HB-cont} and \ref{tab:WB-cont} (for wino-bino and higgsino-bino models), we used
\beq
M_T(W)>720\,{\rm{GeV}}, \quad \MET>500\,{\rm{GeV}}, \quad \frac{p_T(\ell_2)}{p_T(\ell_1)} > 0.2.
\eeq

Note that in all the above tables, NLSP is 1 TeV and LSP is massless.

%%%%%%

%%%%%%%%%%%%%%%%%%%%%%


\begin{thebibliography}{99}
\baselineskip=15pt

%\cite{Wells:2003tf,ArkaniHamed:2004fb,Giudice:2004tc,Wells:2004di,Arvanitaki:2012ps,Altmannshofer:2013lfa,Randall:1998uk,Giudice:1998xp,ArkaniHamed:2012gw,Kahn:2013pfa,Bhattacherjee:2012ed}
\bibitem{Wells:2003tf} 
  J.~D.~Wells,
  ``Implications of supersymmetry breaking with a little hierarchy between gauginos and scalars,'' Proceedings SUSY 2003 (Tucson, Arizona),
  hep-ph/0306127.
  %%CITATION = HEP-PH/0306127;%%

%\cite{ArkaniHamed:2004fb}
\bibitem{ArkaniHamed:2004fb} 
  N.~Arkani-Hamed and S.~Dimopoulos,
  ``Supersymmetric unification without low energy supersymmetry and signatures for fine-tuning at the LHC,''
  JHEP {\bf 0506}, 073 (2005)
  [hep-th/0405159].
  %%CITATION = HEP-TH/0405159;%%

%\cite{Giudice:2004tc}
\bibitem{Giudice:2004tc} 
  G.~F.~Giudice and A.~Romanino,
  ``Split supersymmetry,''
  Nucl.\ Phys.\ B {\bf 699}, 65 (2004)
  [Erratum-ibid.\ B {\bf 706}, 65 (2005)]
  [hep-ph/0406088].
  %%CITATION = HEP-PH/0406088;%%
        
%\cite{Wells:2004di}
\bibitem{Wells:2004di} 
  J.~D.~Wells,
  ``PeV-scale supersymmetry,''
  Phys.\ Rev.\ D {\bf 71}, 015013 (2005)
  [hep-ph/0411041].
  %%CITATION = HEP-PH/0411041;%%
  
  %\cite{Randall:1998uk}
\bibitem{Randall:1998uk} 
  L.~Randall and R.~Sundrum,
  ``Out of this world supersymmetry breaking,''
  Nucl.\ Phys.\ B {\bf 557}, 79 (1999)
  [hep-th/9810155].
  %%CITATION = HEP-TH/9810155;%% 

%\cite{Giudice:1998xp}
\bibitem{Giudice:1998xp} 
  G.~F.~Giudice, M.~A.~Luty, H.~Murayama and R.~Rattazzi,
  ``Gaugino mass without singlets,''
  JHEP {\bf 9812}, 027 (1998)
  [hep-ph/9810442].
  %%CITATION = HEP-PH/9810442;%%
  

%\cite{Bhattacherjee:2012ed}
 \bibitem{Bhattacherjee:2012ed}   
  B.~Bhattacherjee, B.~Feldstein, M.~Ibe, S.~Matsumoto and T.~T.~Yanagida,
  ``Pure Gravity Mediation of Supersymmetry Breaking at the LHC,''
  Phys.\ Rev.\ D {\bf 87}, 015028 (2013)
  [arXiv:1207.5453 [hep-ph]].
  %%CITATION = ARXIV:1207.5453;%%

%\cite{Arvanitaki:2012ps}
\bibitem{Arvanitaki:2012ps} 
  A.~Arvanitaki, N.~Craig, S.~Dimopoulos and G.~Villadoro,
  ``Mini-Split,''
  JHEP {\bf 1302}, 126 (2013)
  [arXiv:1210.0555 [hep-ph]].
  %%CITATION = ARXIV:1210.0555;%%

%\cite{ArkaniHamed:2012gw}
\bibitem{ArkaniHamed:2012gw} 
  N.~Arkani-Hamed, A.~Gupta, D.~E.~Kaplan, N.~Weiner and T.~Zorawski,
  ``Simply Unnatural Supersymmetry,''
  arXiv:1212.6971 [hep-ph].
  %%CITATION = ARXIV:1212.6971;%%

%\cite{Kahn:2013pfa}
\bibitem{Kahn:2013pfa} 
  Y.~Kahn, M.~McCullough and J.~Thaler,
  ``Auxiliary Gauge Mediation: A New Route to Mini-Split Supersymmetry,''
  JHEP {\bf 1311}, 161 (2013)
  [arXiv:1308.3490 [hep-ph]].
  %%CITATION = ARXIV:1308.3490;%%  

%\cite{Altmannshofer:2013lfa}
\bibitem{McKeen:2013dma} 
  D.~McKeen, M.~Pospelov and A.~Ritz,
  ``Electric dipole moment signatures of PeV-scale superpartners,''
  Phys.\ Rev.\ D {\bf 87}, no. 11, 113002 (2013)
  [arXiv:1303.1172 [hep-ph]].
  %%CITATION = ARXIV:1303.1172;%%
%\bibitem{Altmannshofer:2013lfa} 
  W.~Altmannshofer, R.~Harnik and J.~Zupan,
  ``Low Energy Probes of PeV Scale Sfermions,''
  JHEP {\bf 1311}, 202 (2013)
  [arXiv:1308.3653 [hep-ph], arXiv:1308.3653].
  %%CITATION = ARXIV:1308.3653;%%
  %\cite{McKeen:2013dma}
    

      
      
      
      
%\cite{Cohen:2013xda}
\bibitem{Cohen:2013xda} 
  T.~Cohen, T.~Golling, M.~Hance, A.~Henrichs, K.~Howe, J.~Loyal, S.~Padhi and J.~G.~Wacker,
  ``SUSY Simplified Models at 14, 33, and 100 TeV Proton Colliders,''
  arXiv:1311.6480 [hep-ph].
  %%CITATION = ARXIV:1311.6480;%%

%\cite{Jung:2013zya}
\bibitem{Jung:2013zya} 
  S.~Jung and J.~D.~Wells,
  ``Gaugino physics of split supersymmetry spectrum at the LHC and future proton colliders,''
  Phys.\ Rev.\ D {\bf 89}, 075004 (2014)
  [arXiv:1312.1802 [hep-ph]].
  %%CITATION = ARXIV:1312.1802;%%
  
  

%\cite{ArkaniHamed:2006mb}
\bibitem{ArkaniHamed:2006mb} 
  N.~Arkani-Hamed, A.~Delgado and G.~F.~Giudice,
  ``The Well-tempered neutralino,''
  Nucl.\ Phys.\ B {\bf 741}, 108 (2006)
  [hep-ph/0601041].
  %%CITATION = HEP-PH/0601041;%%
  
%\cite{Fan:2013faa}
\bibitem{Fan:2013faa} 
  J.~Fan and M.~Reece,
  ``In Wino Veritas? Indirect Searches Shed Light on Neutralino Dark Matter,''
  JHEP {\bf 1310}, 124 (2013)
  [arXiv:1307.4400 [hep-ph]].
  %%CITATION = ARXIV:1307.4400;%%
  

%\cite{Low:2014cba}
\bibitem{Low:2014cba} 
  M.~Low and L.~-T.~Wang,
  ``Neutralino Dark Matter at 100 TeV,''
  arXiv:1404.0682 [hep-ph].
  %%CITATION = ARXIV:1404.0682;%%      

%\cite{Cirelli:2014dsa}
\bibitem{Cirelli:2014dsa} 
  M.~Cirelli, F.~Sala and M.~Taoso,
  ``Wino-like Minimal Dark Matter and future colliders,''
  arXiv:1407.7058 [hep-ph].
  %%CITATION = ARXIV:1407.7058;%%
  
  
%\cite{Hook:2014rka}
\bibitem{Hook:2014rka} 
  A.~Hook and A.~Katz,
  ``Unbroken $SU(2)$ at a 100 TeV collider,''
  arXiv:1407.2607 [hep-ph].
  %%CITATION = ARXIV:1407.2607;%%


%\cite{Jung:2014bda}
\bibitem{Jung:2014bda} 
  S.~Jung,
  ``Resolving the existence of Higgsinos in the LHC inverse problem,''
  JHEP {\bf 1406}, 111 (2014)
  [arXiv:1404.2691 [hep-ph]].
  %%CITATION = ARXIV:1404.2691;%%

%\cite{Jung:work}
\bibitem{Jung:work} 
  E.~J.~Chun, S.~Jung and P.~Sharma,
  Work in progress.
      
%\cite{Han:2013kza}
\bibitem{Han:2013kza}
  T.~Han, S.~Padhi and S.~Su,
  ``Electroweakinos in the Light of the Higgs Boson,''
  Phys.\ Rev.\ D {\bf 88}, 115010 (2013)
  [arXiv:1309.5966 [hep-ph]].
  %%CITATION = ARXIV:1309.5966;%%





%\cite{Aad:2014nua}
\bibitem{Aad:2014nua}
  G.~Aad {\it et al.}  [ATLAS Collaboration],
  %``Search for direct production of charginos and neutralinos in events with three leptons and missing transverse momentum in $\sqrt{s}$ = 8 TeV $pp$ collisions with the ATLAS detector,''
  arXiv:1402.7029 [hep-ex].
  %%CITATION = ARXIV:1402.7029;%%

%\cite{Aad:2014pda}
\bibitem{Aad:2014pda} 
  G.~Aad {\it et al.}  [ATLAS Collaboration],
  %``Search for supersymmetry at $\sqrt{s}$=8 TeV in final states with jets and two same-sign leptons or three leptons with the ATLAS detector,''
  JHEP {\bf 1406}, 035 (2014)
  [arXiv:1404.2500 [hep-ex]].
  %%CITATION = ARXIV:1404.2500;%%

%\cite{Khachatryan:2014qwa}
\bibitem{Khachatryan:2014qwa}
  V.~Khachatryan {\it et al.}  [CMS Collaboration],
  %``Searches for electroweak production of charginos, neutralinos, and sleptons decaying to leptons and W, Z, and Higgs bosons in pp collisions at 8 TeV,''
  arXiv:1405.7570 [hep-ex].
  %%CITATION = ARXIV:1405.7570;%%

%\cite{ATLAS:2013qla}
\bibitem{ATLAS:2013qla}
  [ATLAS Collaboration],
  %``Search for supersymmetry in events with four or more leptons in 21$\,$fb$^{-1}$ of pp collisions at $\sqrt{s}=8\,$TeV with the ATLAS detector,''
  ATLAS-CONF-2013-036.
  %%CITATION = ATLAS-CONF-2013-036;%%

%\cite{Aad:2014vma}
\bibitem{Aad:2014vma}
  G.~Aad {\it et al.}  [ATLAS Collaboration],
  %``Search for direct production of charginos, neutralinos and sleptons in final states with two leptons and missing transverse momentum in $pp$ collisions at $\sqrt{s} =$ 8 TeV with the ATLAS detector,''
  JHEP {\bf 1405}, 071 (2014)
  [arXiv:1403.5294 [hep-ex]].
  %%CITATION = ARXIV:1403.5294;%%

       

  
%\cite{Baer:2012ts}
\bibitem{Baer:2012ts} 
  H.~Baer, V.~Barger, A.~Lessa, W.~Sreethawong and X.~Tata,
  ``$Wh$ plus missing-$E_T$ signature from gaugino pair production at the LHC,''
  Phys.\ Rev.\ D {\bf 85}, 055022 (2012)
  [arXiv:1201.2949]. \\
  %%CITATION = ARXIV:1201.2949;%%
  %\cite{Howe:2012xe}
%\bibitem{Howe:2012xe} 
  K.~Howe and P.~Saraswat,
  ``Excess Higgs Production in Neutralino Decays,''
  JHEP {\bf 1210}, 065 (2012)
  [arXiv:1208.1542 [hep-ph]]. \\
  %%CITATION = ARXIV:1208.1542;%%
  %\cite{Arbey:2012fa}
%\bibitem{Arbey:2012fa} 
  A.~Arbey, M.~Battaglia and F.~Mahmoudi,
  ``Higgs Production in Neutralino Decays in the MSSM - The LHC and a Future $e^+e^-$ Collider,''
  arXiv:1212.6865 [hep-ph].
  %%CITATION = ARXIV:1212.6865;%%
  %\cite{Bharucha:2013epa}

%\cite{Gaunt:2010pi}
\bibitem{Gaunt:2010pi} 
  J.~R.~Gaunt, C.~-H.~Kom, A.~Kulesza and W.~J.~Stirling,
  ``Same-sign W pair production as a probe of double parton scattering at the LHC,''
  Eur.\ Phys.\ J.\ C {\bf 69}, 53 (2010)
  [arXiv:1003.3953 [hep-ph]].
  %%CITATION = ARXIV:1003.3953;%%
  
%\cite{ATLAS:2012mn}
\bibitem{ATLAS:2012mn} 
  G.~Aad {\it et al.}  [ATLAS Collaboration],
  %``Search for anomalous production of prompt like-sign lepton pairs at $\sqrt{s}=7$ TeV with the ATLAS detector,''
  JHEP {\bf 1212}, 007 (2012)
  [arXiv:1210.4538 [hep-ex]].
  %%CITATION = ARXIV:1210.4538;%%
  




%\cite{Alwall:2011uj}
\bibitem{Alwall:2011uj} 
  J.~Alwall, M.~Herquet, F.~Maltoni, O.~Mattelaer and T.~Stelzer,
  ``MadGraph 5 : Going Beyond,''
  JHEP {\bf 1106}, 128 (2011)
  [arXiv:1106.0522 [hep-ph]].
  %%CITATION = ARXIV:1106.0522;%%

%\cite{Sjostrand:2006za}
\bibitem{Sjostrand:2006za} 
  T.~Sjostrand, S.~Mrenna and P.~Z.~Skands,
  ``PYTHIA 6.4 Physics and Manual,''
  JHEP {\bf 0605}, 026 (2006)
  [hep-ph/0603175].
  %%CITATION = HEP-PH/0603175;%%

%\cite{Mangano:2006rw}
\bibitem{Mangano:2006rw} 
  M.~L.~Mangano, M.~Moretti, F.~Piccinini and M.~Treccani,
  ``Matching matrix elements and shower evolution for top-quark production in hadronic collisions,''
  JHEP {\bf 0701}, 013 (2007)
  [hep-ph/0611129].
  %%CITATION = HEP-PH/0611129;%%
  %\cite{Anderson:2013kxz}
  
  %\cite{Anderson:2013kxz}
\bibitem{Anderson:2013kxz} 
  J.~Anderson, A.~Avetisyan, R.~Brock, S.~Chekanov, T.~Cohen, N.~Dhingra, J.~Dolen and J.~Hirschauer {\it et al.},
  ``Snowmass Energy Frontier Simulations,''
  arXiv:1309.1057 [hep-ex].
  %%CITATION = ARXIV:1309.1057;%%
  %32 citations counted in INSPIRE as of 13 Oct 2014

%\cite{Campbell:1999ah}
\bibitem{Campbell:1999ah} 
  J.~M.~Campbell and R.~K.~Ellis,
  ``An Update on vector boson pair production at hadron colliders,''
  Phys.\ Rev.\ D {\bf 60}, 113006 (1999)
  [hep-ph/9905386].
  %%CITATION = HEP-PH/9905386;%%
  
  %\cite{Cacciari:2008gp}
\bibitem{Cacciari:2008gp} 
  M.~Cacciari, G.~P.~Salam and G.~Soyez,
  ``The Anti-k(t) jet clustering algorithm,''
  JHEP {\bf 0804}, 063 (2008)
  [arXiv:0802.1189 [hep-ph]].
  %%CITATION = ARXIV:0802.1189;%%

    %\cite{Cacciari:2011ma}
\bibitem{Cacciari:2011ma} 
  M.~Cacciari, G.~P.~Salam and G.~Soyez,
  ``FastJet User Manual,''
  Eur.\ Phys.\ J.\ C {\bf 72}, 1896 (2012)
  [arXiv:1111.6097 [hep-ph]].
  %%CITATION = ARXIV:1111.6097;%%
  
  %\cite{Gori:2013ala}
\bibitem{Gori:2013ala} 
  S.~Gori, S.~Jung and L.~T.~Wang,
  ``Cornering electroweakinos at the LHC,''
  JHEP {\bf 1310}, 191 (2013)
  [arXiv:1307.5952 [hep-ph]].
  %%CITATION = ARXIV:1307.5952;%%
  %12 citations counted in INSPIRE as of 04 Sep 2014   
  
%\cite{Aad:2011mk}
\bibitem{Aad:2011mk} 
  G.~Aad {\it et al.}  [ATLAS Collaboration],
  %``Electron performance measurements with the ATLAS detector using the 2010 LHC proton-proton collision data,''
  Eur.\ Phys.\ J.\ C {\bf 72}, 1909 (2012)
  [arXiv:1110.3174 [hep-ex]].
  %%CITATION = ARXIV:1110.3174;%%
  %539 citations counted in INSPIRE as of 13 Sep 2014
  
  %\cite{Aad:2009wy}
\bibitem{Aad:2009wy} 
  G.~Aad {\it et al.}  [ATLAS Collaboration],
  %``Expected Performance of the ATLAS Experiment - Detector, Trigger and Physics,''
  arXiv:0901.0512 [hep-ex].
  %%CITATION = ARXIV:0901.0512;%%
  %1599 citations counted in INSPIRE as of 13 Sep 2014
  
  %\cite{Chatrchyan:2013iaa}
\bibitem{Chatrchyan:2013iaa} 
  S.~Chatrchyan {\it et al.}  [CMS Collaboration],
  %``Measurement of Higgs boson production and properties in the WW decay channel with leptonic final states,''
  JHEP {\bf 1401}, 096 (2014)
  [arXiv:1312.1129 [hep-ex]].
  %%CITATION = ARXIV:1312.1129;%%
  
    
  %\cite{Barenboim:2014kka}
\bibitem{Barenboim:2014kka} 
  G.~Barenboim, E.~J.~Chun, S.~Jung and W.~I.~Park,
  ``Implications of the axino LSP on the naturalness,''
  arXiv:1407.1218 [hep-ph].
  %%CITATION = ARXIV:1407.1218;%%
  
  %\cite{Cohen:2013ama}
\bibitem{Cohen:2013ama} 
  T.~Cohen, M.~Lisanti, A.~Pierce and T.~R.~Slatyer,
  ``Wino Dark Matter Under Siege,''
  JCAP {\bf 1310}, 061 (2013)
  [arXiv:1307.4082].
  %%CITATION = ARXIV:1307.4082;%%

  
%\cite{Gunion:1987yh}
\bibitem{Gunion:1987yh} 
  J.~F.~Gunion and H.~E.~Haber,
  ``Two-body Decays of Neutralinos and Charginos,''
  Phys.\ Rev.\ D {\bf 37}, 2515 (1988).
  %%CITATION = PHRVA,D37,2515;%%
  
  %\cite{Choi:2007ka}
\bibitem{Choi:2007ka} 
  K.~Choi and H.~P.~Nilles,
  ``The Gaugino code,''
  JHEP {\bf 0704}, 006 (2007)
  [hep-ph/0702146].
  %%CITATION = HEP-PH/0702146;%%

%\cite{Gupta:2012gu}
\bibitem{Gupta:2012gu} 
  A.~Gupta, D.~E.~Kaplan and T.~Zorawski,
  ``Gaugomaly Mediation Revisited,''
  arXiv:1212.6969 [hep-ph].
  %%CITATION = ARXIV:1212.6969;%%
  
%\cite{ATLAS:2013hta}
\bibitem{ATLAS:2013hta} 
  [ATLAS Collaboration],
  %``Physics at a High-Luminosity LHC with ATLAS,''
  arXiv:1307.7292 [hep-ex].
  %%CITATION = ARXIV:1307.7292;%%

%\cite{CMS:2013xfa}
\bibitem{CMS:2013xfa} 
  [CMS Collaboration],
  %``Projected Performance of an Upgraded CMS Detector at the LHC and HL-LHC: Contribution to the Snowmass Process,''
  arXiv:1307.7135.
  %%CITATION = ARXIV:1307.7135;%%

%\cite{Schwaller:2013baa}  
\bibitem{Schwaller:2013baa} 
  P.~Schwaller and J.~Zurita,
  %``Compressed electroweakino spectra at the LHC,''
  JHEP {\bf 1403}, 060 (2014)
  [arXiv:1312.7350 [hep-ph]].
  %%CITATION = ARXIV:1312.7350;%%
  %\cite{Han:2014kaa}
%\bibitem{Han:2014kaa} 
  Z.~Han, G.~D.~Kribs, A.~Martin and A.~Menon,
  ``Hunting quasidegenerate Higgsinos,''
  Phys.\ Rev.\ D {\bf 89}, 075007 (2014)
  [arXiv:1401.1235 [hep-ph]].
  %%CITATION = ARXIV:1401.1235;%%
        
  %\cite{Acharya:2014pua}
\bibitem{Acharya:2014pua} 
  B.~S.~Acharya, K.~Bozek, C.~Pongkitivanichkul and K.~Sakurai,
  ``Prospects for observing charginos and neutralinos at a 100 TeV proton-proton collider,''
  arXiv:1410.1532 [hep-ph].
  %%CITATION = ARXIV:1410.1532;%%
  
%\cite{colliderreach}
\bibitem{colliderreach} 
Collier Reach (beta). http://collider-reach.web.cern.ch/



  
  
\end{thebibliography}
\end{document}